\documentclass[%
reprint,
superscriptaddress,
showpacs,
preprintnumbers,
amsmath,
amssymb,
aps,
prd,
]{revtex4-2}

\usepackage{graphicx}
\usepackage{dcolumn}
\usepackage{bm}

\begin{document}

\title{Fermion scattering on topological solitons in the $\mathbb{CP}^{N-1}$ model}

\author{A.~Yu.~Loginov}

\email{a.yu.loginov@tusur.ru}

\affiliation{Laboratory of Applied Mathematics and Theoretical Physics, Tomsk State
             University of Control Systems and Radioelectronics, 634050 Tomsk, Russia}


\date{\today}

\begin{abstract}
The scattering of  Dirac  fermions  in  the  background  fields  of topological
solitons of the $(2+1)$-dimensional $\mathbb{CP}^{N-1}$  model is studied using
analytical and numerical methods.
It is  shown  that  the  exact solutions  for  fermionic  wave functions can be
expressed in terms of the confluent Heun functions.
The question of the existence  of  bound  states for the fermion-soliton system
is then investigated.
General formulae  describing  fermion  scattering  are obtained, and a symmetry
property for the partial phase shifts is derived.
The  amplitudes  and   cross-sections   of  the  fermion-soliton scattering are
obtained in an analytical form within the framework  of the Born approximation,
and the symmetry properties  and  asymptotic  forms  of the Born amplitudes are
investigated.
The dependences of the first few partial phase shifts  on  the fermion momentum
are obtained  by   numerical   methods,   and   some  of  their  properties are
investigated and discussed.
\end{abstract}

\maketitle

\section{\label{sec:I} Introduction}

A number of $(2 + 1)$-dimensional  field  models  admit the existence of planar
topological  solitons  \cite{Manton,  E_Weinberg,  Zakrzewski},  which  play an
important role in field  theory, high-energy physics, condensed matter physics,
cosmology, and hydrodynamics.
The vortices of the effective theory of superconductivity \cite{abr} and of the
$\left(2+1\right)$-dimensional Abelian  Higgs model \cite{nielsen} are probably
the most important topological solitons of this type.
The next most important is the topological soliton of the $(2 + 1)$-dimensional
nonlinear $O\left(3\right)$ sigma model \cite{BP}.
One feature of the soliton solutions of the nonlinear $O(3)$ sigma model is the
presence of an arbitrary parameter determining their spatial size.
This is because the static energy functional of the $(2 +1)$-dimensional $O(3)$
sigma model is invariant under scale transformations.

Nonlinear sigma models can  also  be  formulated  for  orthogonal groups $O(N)$
with $N \ge 4$, but unlike the $O(3)$ sigma model, these models have no soliton
solutions.
However, there  is  another  family  of  nonlinear  scalar  field  models whose
properties are  similar  to  those of the nonlinear $O(N)$ sigma models in many
respects, but which  have topological soliton solutions for an arbitrary number
of fields.
These are the so-called $\mathbb{CP}^{N - 1}$ models \cite{cremmer, eichenherr,
 per1, per2}.
For $N=2$, the $\mathbb{CP}^{N-1}$  model is reduced to the $O(3)$ sigma model,
but for $N \ge 3$, the $\mathbb{CP}^{N - 1}$  model  is a better generalization
than the $O(N + 1)$  sigma  model,  as  it  continues to have soliton solutions
\cite{dadda, witten} even in this case.

Since their appearance in the late 1970s, the $\mathbb{CP}^{N - 1}$ models have
consistently  attracted  interest,  primarily   based  on  the  fact  that  the
two-dimensional $\mathbb{CP}^{N-1}$ models are an useful instrument for studying
nonperturbative effects in the four-dimensional Yang-Mills models.
The two-dimensional $\mathbb{CP}^{N - 1}$  models  share many common properties
with four-dimensional Yang-Mills models, including  conformal invariance at the
classical level, asymptotic freedom in the ultraviolet regime \cite{polyakov2},
strong coupling  in  the  infrared  regime,  and the existence of a topological
term and instantons \cite{dadda, witten} resulting  in  a  complex structure of
the vacuum at the quantum level.
The lower dimensionality of the  $\mathbb{CP}^{N - 1}$  models  facilitates the
analysis of nonperturbative effects in the strong coupling regime, compared  to
the more complex four-dimensional Yang-Mills models.
In addition,  two-dimensional $\mathbb{CP}^{N - 1}$ models can be considered as
effective field theories  describing low-energy dynamics on the  world sheet of
non-Abelian  vortex  strings  in  a  class  of  four-dimensional gauge theories
\cite{hanany_2003,  auzzi_2003,  shifman_2003,  hanany_2004,  shif_yung, tong}.
The $\mathbb{CP}^{N-1}$ models also  have interesting applications in the field
of   condensed   matter    physics   \cite{Tsvelik},    and    particularly  in
(anti)ferromagnetism, the Hall effect, and the Kondo  effect.
They also find application in the study of the sphaleron-induced fermion number
violation at high temperature \cite{mottola_1989}.

The $\mathbb{CP}^{N - 1}$  model  can  be  extended to include fermionic matter
fields.
This can be achieved either by a supersymmetric extension of the $\mathbb{CP}^{
N-1}$ model or by  minimal coupling  between  fermionic  fields and a composite
gauge field of the $\mathbb{CP}^{N - 1}$  model  (the  so-called  minimal model
\cite{abdalla_1982}).
The supersymmetric extension of the $\mathbb{CP}^{N-1}$ model involves Majorana
fermion fields that satisfy nontrivial constraints,  whereas  the minimal model
deals with unconstrained Dirac fermion fields.
In the present paper, we investigate  a  fermion-soliton  system in the minimal
model within the background field approximation.
In particular, we find  that the  fermionic  wave  functions  are  expressed in
terms of the confluent Heun functions,  and that the fermion-soliton system has
no bound states.
The results obtained here can be used to describe the  interaction  of fermions
with two-dimensional or thread-like  three-dimensional   topological defects in
condensed matter physics.
We note  that  it  was  stated  in  Ref.~\cite{birkandan_2017}  that  the  wave
functions of  a  Dirac   fermion   minimally  coupled  to  the  two-dimensional
$\mathbb{CP}^{1}$ model  can  be  expressed  in  terms  of  the  confluent Heun
functions.
Furthermore, it  was  shown in Refs.~\cite{loginov_epjc_2022, loginov_npb_2022}
that the fermion  scattering  on  a  one-dimensional  kink  or  Q-ball can also
be described in terms of  Heun-type functions.

This  paper  is  structured as follows.
In Sec.~\ref{sec:II}, we describe briefly  the  Lagrangian,   symmetries, field
equations, and  topological solitons  of the  $\mathbb{CP}^{N-1}$ model.
In Sec.~\ref{sec:III},  we   study   the   fermion-soliton   scattering  in the
background field approximation.
We show that for $\mathbb{CP}^{N-1}$ solitons with winding numbers $n = \pm 1$,
the fermionic wave functions can  be  expressed  in terms of the confluent Heun
functions.
We also consider the question  of  the  existence of bound fermionic states for
these solitons, and  establish  a  symmetry  property for partial phase shifts.
In Sec.~\ref{sec:IV},  we  give  an   analytical   description  of  the fermion
scattering within the framework of the Born approximation.
In Sec.~\ref{sec:V}, we present numerical results  for  the  first  few partial
phase shifts, and compare  the   exact  results with those obtained in the Born
approximation.
In the final section, we briefly  summarize the results obtained in the present
work.

Throughout the paper, the natural units $c = 1$ and $\hbar = 1$ are used.

\section{\label{sec:II}  Lagrangian, field  equations, and topological solitons
of the model}

The Lagrangian density of the $\mathbb{CP}^{N - 1}$ model minimally interacting
with fermionic fields has the form
\begin{equation}
\mathcal{L}=g^{-1}\left( D_{\mu }n_{a}\right)^{\ast }D^{\mu }n_{a}+i
\bar{\psi}_{a}\gamma^{\mu}D_{\mu}\psi_{a}-M\bar{\psi}_{a}\psi_{a}, \label{II:1}
\end{equation}
where  $n_{a}$ are complex scalar  fields, $\psi_{a}$ are fermionic fields, $g$
is a coupling constant, and the index $a$ runs from  one  to  $N$.
The complex  scalar  fields  $n_{a}$   satisfy   the   normalization  condition
$n_{a}^{\ast }n_{a} = 1$,  where  summation  over  repeated indices is implied.
In Eq.~(\ref{II:1}), the covariant derivatives of fields are
\begin{subequations}                                               \label{II:2}
\begin{eqnarray}
D_{\mu}n_{a} &=&\partial_{\mu}n_{a} + i A_{\mu}n_{a},             \label{II:2a}
 \\
D_{\mu}\psi_{a} &=&\partial_{\mu}\psi_{a} + i A_{\mu}\psi_{a},    \label{II:2b}
\end{eqnarray}
\end{subequations}
where $A_{\mu}$ is a vector gauge field.

By varying  the  action $S = \int \mathcal{L}d^{2}xdt$  in  the fields $n_{a}$,
$\bar{\psi}_{a}$,  and  $A_{\mu}$,  and  taking  into  account  the  constraint
$n_{a}^{\ast}n_{a} = 1$  by  means of the Lagrange multiplier method, we obtain
the field equations for the minimal $\mathbb{CP}^{N - 1}$ model:
\begin{eqnarray}
D_{\mu}D^{\mu}n_{a}-\left(n_{b}^{\ast}D_{\mu}D^{\mu}n_{b}\right) n_{a}
&=&0,                                                             \label{II:3a}
\\
\left(i\gamma^{\mu} D_{\mu} - M\right)\psi_{a} &=&0,              \label{II:3b}
\\
A_{\mu }-in_{a}^{\ast }\partial _{\mu }n_{a}-\frac{g}{2}\bar{\psi}
_{a}\gamma_{\mu }\psi_{a} &=&0.                                   \label{II:3c}
\end{eqnarray}
It follows  from  Eq.~(\ref{II:3c})  that  the  gauge  field $A_{\mu}$ is not a
dynamic one, and is expressed in terms of the fields $n_{a}$ and $\psi_{a}$.

The  minimal $\mathbb{CP}^{N - 1}$  model  possesses  a  number  of symmetries.
The invariance of model (\ref{II:1})  under  the  global $U(N)$ transformations
$n_{a}\rightarrow U_{ab} n_{b},\,\psi_{a} \rightarrow U_{ab} \psi_{b}$  results
in the existence of the Noether current, which is a vector field with values in
complex anti-Hermitian $(N\! \times\! N)$-matrices, with matrix entries
\begin{equation}
\left(j_{\mu }\right)_{ab}=g^{-1}\left[n_{a}\left(D_{\mu}n_{b}\right)
^{\ast }-\left( D_{\mu }n_{a}\right) n_{b}^{\ast }\right] + i \bar{\psi}
_{a}\gamma _{\mu }\psi _{b}.                                       \label{II:4}
\end{equation}
In addition, the model is also invariant under the local $U(1)$ transformations
\begin{subequations}                                               \label{II:5}
\begin{eqnarray}
n_{a}\left( x\right)  &\rightarrow &e^{i\Lambda \left( x\right) }n_{a}\left(
x\right),                                                         \label{II:5a}
\\
\psi _{a}\left( x\right)  &\rightarrow &e^{i\Lambda \left( x\right) }\psi
_{a}\left( x\right),                                              \label{II:5b}
\\
A_{\mu }\left( x\right)  &\rightarrow &A_{\mu }\left( x\right)
+ i e^{-i\Lambda \left( x\right) }\partial_{\mu }e^{i\Lambda \left(x\right)}.
                                                                  \label{II:5c}
\end{eqnarray}
\end{subequations}
The corresponding gauge current is the  trace  of  the matrix-valued current in
Eq.~(\ref{II:4}).

A characteristic feature of the $\mathbb{CP}^{N-1}$ models is that they possess
localized solutions \cite{dadda, witten},  which can be  interpreted  either as
instantons in  the  two-dimensional  Euclidean  case  or  as static topological
solitons in the $(2 + 1)$-dimensional case.
All of these solutions can be obtained in analytical form.
In   particular, the  $\mathbb{Z}_{\left\vert n \right\vert}$ symmetric soliton
solution of the $(2+1)$-dimensional $\mathbb{CP}^{N-1}$ model can be written as
\begin{equation}
\mathbf{n}\left( \rho ,\theta \right) =\frac{\lambda ^{\left\vert
n\right\vert }\mathbf{u}+\rho ^{\left\vert n\right\vert }e^{in\theta }
\mathbf{v}}{\left( \lambda ^{2\left\vert n\right\vert }+\rho ^{2\left\vert
n\right\vert }\right) ^{1/2}},                                     \label{II:6}
\end{equation}
where $n$ is a non-zero  integer,  $\rho$  and  $\theta$ are polar coordinates,
$\mathbf{u} = \left(1,0,\ldots,0\right)$  and  $\mathbf{v}=\left( 0, 0, \ldots,
1 \right)$ are orthonormal $N$-dimensional vectors, and  $\lambda$ is a scaling
parameter, which determines the effective size of the soliton.
The gauge field $A_{\mu}$ that corresponds to solution (\ref{II:6}) is
\begin{equation}
A_{\mu} = n\frac{\rho ^{2\left(\left\vert
n\right\vert -1\right) }}{\rho ^{2\left\vert n\right\vert }+\lambda
^{2\left\vert n\right\vert}}\left(0, y, -x\right),                 \label{II:7}
\end{equation}
where we factor out the common factor of the covariant  components of the gauge
field.

Eq.~(\ref{II:6}) tells us that $\mathbf{n}\rightarrow e^{i n \theta}\mathbf{v}$
as $\rho \rightarrow \infty$, and, consequently, solution (\ref{II:6}) tends to
the same element of the complex  projective space $\mathbb{CP}^{N - 1}$ in this
limit.
It  follows  that  field   configurations   (\ref{II:6})   are  maps  from  the
compactified  plane  (which  is  topologically  equivalent  to  the  two-sphere
$S^{2}$) to $\mathbb{CP}^{N - 1}$.
Since the  second  homotopic  group  $\pi_{2}\bigl(\mathbb{CP}^{N - 1} \bigr) =
\mathbb{Z}$, field  configurations (\ref{II:6}) can  be labelled  by an integer
$Q$ called the winding number, as explicitly given in Ref.~\cite{dadda}:
\begin{equation}
Q=-\frac{1}{2\pi}\int d^{2}x\,\epsilon_{ij}\partial_{i}A_{j} =
  -\frac{1}{2\pi}\int\limits_{S^{1}}A_{i}dx^{i},                   \label{II:8}
\end{equation}
where  $\epsilon_{i j}$  is   the   two-dimensional  antisymmetric  tensor  and
$\epsilon_{1 2} = 1$.
It can easily be shown that  for  soliton  solutions  (\ref{II:6}), the winding
number $Q  =  n$,  meaning  that  they  are  stable  to  transition  into field
configurations belonging to the topologically trivial sector with $n = 0$.

In contrast to the  usual  solutions  satisfying  field equations of the second
order, solution (\ref{II:6}) also  satisfies  the equations of the first order,
\begin{equation}
D_{i}\mathbf{n}\pm i\epsilon_{ij}D_{j}\mathbf{n} = 0.             \label{II:9}
\end{equation}
Depending on  the  sign, this  is  called the self-duality or anti-self-duality
condition.
It can be shown \cite{dadda} that the energy of any solution in the topological
sector with $Q = n$ satisfies the inequality
\begin{equation}
E \ge 2 \pi \left\vert n \right\vert g^{-1},                      \label{II:10}
\end{equation}
and  that  saturation  of  this  inequality  is  possible  only  for  solutions
satisfying Eq.~(\ref{II:9}).
Note that for $N > 2$, the  $\mathbb{CP}^{N - 1}$  models  also have additional
solutions \cite{dz1, dz2, dz3} that  do  not  satisfy  Eq.~(\ref{II:9}) and are
only unstable saddle points of the energy functional.

The energy density of soliton solution (\ref{II:6}) is
\begin{eqnarray}
\mathcal{E} &=&g^{-1}\left[\left(D_{0}n_{a}\right)^{\ast }D_{0}n_{a}+\left(
D_{i}n_{a}\right) ^{\ast }D_{i}n_{a}\right]                           \nonumber
\\
&=&2n^{2}\lambda ^{2}g^{-1}\frac{\left( \lambda \rho \right)^{2\left(
\left\vert n\right\vert-1\right)}}{\left(\rho^{2\left\vert n\right\vert
}+\lambda^{2\left\vert n\right\vert }\right)^{2}},                \label{II:11}
\end{eqnarray}
and the soliton energy
\begin{equation}
E_{\text{s}}=2\pi\int\limits_{0}^{\infty}\mathcal{E}\left(\rho\right) d\rho =
2 \pi \left\vert n \right\vert g^{-1} = 2 \pi \left\vert Q \right\vert g^{-1}.
                                                                  \label{II:12}
\end{equation}
We  see  that  the  soliton  energy  saturates  inequality  (\ref{II:10}),  and
therefore is the absolute minimum in the topological sector with a given $Q$.
The soliton energy does not depend on the scaling parameter $\lambda$ since the
bosonic part of  the  action  of  model (\ref{II:1})  is  invariant under scale
transformations $\mathbf{x} \rightarrow a \mathbf{x}$.

\section{\label{sec:III}   Fermions in the background field of a $\mathbb{CP}^{
N-1}$ soliton}

We  consider  fermion  scattering   on  the  topological  $\mathbb{CP}^{N - 1}$
soliton within the background field approximation, i.e., we neglect the fermion
backreaction on soliton field configuration (\ref{II:6}).
Eq.~(\ref{II:3c}) tells  us  that  the  gauge  field $A_{\mu} =  i n_{a}^{\ast}
\partial_{\mu} n_{a} + 2^{-1} g \bar{\psi}_{a}\gamma_{\mu} \psi_{a}$.
The background  field  approximation  involves  neglecting  the  fermionic term
$2^{-1} g \bar{\psi}_{a} \gamma_{\mu}\psi_{a}$  in  comparison with the bosonic
term $i n_{a}^{\ast}\partial_{\mu} n_{a}$.
In this case, the gauge field $A_{\mu} = in_{a}^{\ast}\partial_{\mu}n_{a}$, and
it follows from  Eqs.~(\ref{II:2})  and  (\ref{II:3a}) that there is no fermion
backreaction on the soliton field.
An analysis shows that this approximation is possible under the condition
\begin{equation}
g \ll \left\vert n\right\vert \lambda^{-1}\varrho _{F}^{-1},      \label{III:I}
\end{equation}
where $\varrho_{F}$ is the two-dimensional density of incident fermions.

The presence of the  fermionic  part  in  the  gauge field $A_{\mu}$ causes the
Dirac equation (\ref{II:3b}) to be nonlinear (cubic) in fermionic fields, which
does not allow us to obtain an analytical solution.
In order to avoid this, we  must  neglect this nonlinear cubic term compared to
the linear Dirac mass term in Eq.~(\ref{II:3b}).
This is possible if the condition
\begin{equation}
g \ll M \varrho _{F}^{-1}                                        \label{III:II}
\end{equation}
is satisfied, where $M$ is the fermion mass.

We see that conditions (\ref{III:I}) and (\ref{III:II}) can always be fulfilled
if the density of incident fermions is sufficiently low.
From the viewpoint of QFT, however, we are  talking  about  the scattering of a
fermion of mass $M$ on a  $\mathbb{CP}^{N-1}$ soliton of mass $M_{\text{s}} = 2
\pi \left\vert n \right\vert g^{-1}$.
To allow us to neglect the recoil of the $\mathbb{CP}^{N-1}$ soliton in fermion
scattering, the  mass  $M_{\text{s}}$  must  be  much  larger  than  the energy
$\varepsilon$ of the incident fermion, which leads to the condition
\begin{equation}
g\ll 2\pi \left\vert n\right\vert \varepsilon ^{-1} <
2\pi \left\vert n\right\vert M^{-1}.                            \label{III:III}
\end{equation}

Conditions (\ref{III:I}), (\ref{III:II}), and (\ref{III:III}) do not contradict
each other, and  can  always  be  satisfied  if  the  coupling  constant $g$ is
sufficiently small.
In this case, the Dirac equation (\ref{II:3b}) can be written in the Hamiltonian
form
\begin{equation}
i \partial_{t}\psi_{a} = H\psi_{a},                               \label{III:1}
\end{equation}
where the Hamiltonian
\begin{equation}
H = -i \left(\partial_{k} - n_{b}^{\ast}\partial_{k}n_{b}\right)\alpha^{k}
+ \beta M,                                                         \label{III:2}
\end{equation}
the matrices $\alpha^{i} = \gamma^{0}\gamma^{i}$  and $\beta = \gamma^{0}$, and
the Dirac matrices
\begin{equation}
\gamma ^{0}=  \sigma_{3},\,
\gamma ^{1}=-i\sigma_{1},\,
\gamma ^{2}=-i\sigma_{2}.                                         \label{III:3}
\end{equation}
Note that all components $\psi_{a}$ of the fermionic multiplet $\left(\psi_{1},
\ldots, \psi_{N}\right)$ satisfy the same equation (\ref{III:1}).

We now discuss the  symmetry  properties  of  the  Dirac equation (\ref{III:1})
under discrete transformations.
Let $\psi(t,\mathbf{x})$  be  a  solution  to  the Dirac equation (\ref{III:1})
in the background field of $\mathbb{CP}^{N - 1}$ soliton (\ref{II:6}).
It can easily be shown that in this case,
\begin{equation}
\psi^{C}\left(t,\mathbf{x}\right) =\sigma_{1}\psi^{\ast}\left(
t,\mathbf{x}\right),                                             \label{III:3a}
\end{equation}
\begin{equation}
\psi^{P}\left(t,\mathbf{x}\right)=\sigma_{3}\psi \left(t,
-\mathbf{x}\right),                                              \label{III:3b}
\end{equation}
and
\begin{equation}
\psi^{\Pi_{2}T}\left( t,x,y\right)=\psi^{\ast}\left(
-t,x,-y\right)                                                   \label{III:3c}
\end{equation}
are also solutions to this equation.
Solutions (\ref{III:3a}), (\ref{III:3b}), and  (\ref{III:3c}) are obtained from
the original  solution  $\psi(t,\mathbf{x})$  by  means  of  the $C$, $P$,  and
combined $\Pi_{2}T$ transformations,  respectively,  where the symbol $\Pi_{2}$
denotes the operation of coordinate reflection about the $Ox_{1}$ axis.

\subsection{\label{subsec:IIIA}  Exact fermionic wave functions}

It can easily be shown  that  the  Hamiltonian  (\ref{III:2}) commutes with the
angular momentum operator
\begin{equation}
J_{3} = -i\partial_{\theta} + \sigma_{3}/2.                       \label{III:4}
\end{equation}
The presence of the conserved angular  momentum $J_{3}$ is due to the fact that
according  to  Eq.~(\ref{II:7}),  the  vector  field  $A_{\mu} = i n_{a}^{\ast}
\partial_{\mu }n_{a}$  in  the  Hamiltonian  (\ref{III:2})  is  invariant under
in-plane rotations. 
The common  eigenfunctions  of  the  operators  $H$  and  $J_{3}$ have the form
\begin{equation}
\psi_{ m}=\left(
\begin{array}{c}
e^{i\left( m-1/2\right) \theta }f\left( \rho \right)  \\
e^{i\left( m+1/2\right) \theta }g\left( \rho \right)
\end{array}%
\right) e^{-i\varepsilon t},                                      \label{III:5}
\end{equation}
where  $\varepsilon$   and   $m$  are  the  eigenvalues  of  $H$  and  $J_{3}$,
respectively.

By substituting Eq.~(\ref{III:5}) into Eq.~(\ref{III:1}), we obtain a system of
first-order  differential  equations  for  the  radial  functions $f(\rho)$ and
$g(\rho)$
\begin{align}
& f^{\prime }\left( \rho \right) = \rho ^{-1}\left( A_{mn}\left( \rho
\right) -1/2\right) f\left( \rho \right) +\left(M\!+\varepsilon \right)
g\left( \rho \right),                                            \label{III:6a}
\end{align}
\begin{align}
& g^{\prime }\left( \rho \right) = \left(M\!-\varepsilon \right) f\left( \rho
\right) -\rho ^{-1}\left( A_{mn}\left( \rho \right) +1/2\right) g\left( \rho
\right),                                                         \label{III:6b}
\end{align}
where
\begin{equation}
A_{mn}\left( \rho \right) =m-n\frac{\rho^{2\left\vert n\right\vert}}{
\lambda^{2\left\vert n\right\vert }+\rho^{2\left\vert n\right\vert}}.
                                                                  \label{III:7}
\end{equation}
The system  of  differential  equations  (\ref{III:6a})  and  (\ref{III:6b}) is
equivalent to the second-order differential equation
\begin{align}
& f^{\prime \prime}\left(\rho\right)+\rho^{-1}f^{\prime}\left(\rho
\right) +\Bigl[ k^{2}-\rho ^{-2} \Bigr.                       \nonumber
\\
& \times \! \left. \left( 1/2-A_{mn}\left( \rho \right) \right) ^{2}-\rho
^{-1}A_{mn}^{\prime}\left(\rho\right)\right]f\left(\rho\right) = 0,
                                                                  \label{III:8}
\end{align}
where $k^{2} = \varepsilon^{2} - M^{2}$, taken  together  with the differential
relation 
\begin{equation}
g\left( \rho \right) =\left[ \rho ^{-1}\left( 1/2-A_{mn}\left( \rho \right)
\right) f\left( \rho \right) +f^{\prime }\left( \rho \right) \right] \left(
M\!+\varepsilon\right)^{-1}.                                      \label{III:9}
\end{equation}
The substitutions $f\rightarrow g,\,g\rightarrow f,\,A_{mn}\rightarrow -A_{mn},
\,\varepsilon \rightarrow -\varepsilon$ in Eqs.~(\ref{III:8}) and (\ref{III:9})
lead to the second-order differential equation  and  differential relation for
the radial functions $g(\rho)$ and $f(\rho)$, respectively.

From Eq.~(\ref{III:7}), it follows that  the  functions $A_{mn}$ depend only on
the dimensionless   combination $\tau = -\rho^{2}/\lambda^{2}$, which therefore
plays the role of a natural independent variable.
In terms of this new variable $\tau$, Eq.~(\ref{III:8}) takes the form
\begin{align}
&f^{\prime \prime }\left( \tau \right) +\tau ^{-1}f^{\prime }\left( \tau
\right) -2^{-2}\tau ^{-1}\Bigl[ k^{2}\lambda ^{2}+\tau ^{-1}\Bigr. \nonumber
\\
&\times \! \left. \left( 1/2-A_{mn}\left( \tau \right) \right)
^{2}+2A_{mn}^{\prime }\left( \tau \right) \right] f\left( \tau \right) = 0,
                                                                 \label{III:10}
\end{align}
where
\begin{equation}
A_{mn}\left( \tau \right)=m-n\frac{\tau^{\left\vert n\right\vert}}{\left(
-1\right)^{\left\vert n\right\vert}+\tau^{\left\vert n\right\vert}}.
                                                                 \label{III:11}
\end{equation}
It follows from Eq.~(\ref{III:11}) that $A_{mn}(\tau)$ has first-order poles at
the points
\begin{equation}
\tau _{k}=e^{i\frac{\pi }{\left\vert n\right\vert }\left( 2k+\left\vert
n\right\vert+1\right)},\;k=0,\ldots ,\left\vert n\right\vert-1. \label{III:11b}
\end{equation}
The  point  $\tau  =  0$   and   the   $\left\vert  n  \right\vert$  points  of
Eq.~(\ref{III:11b}) are  the  regular  singular points of differential equation
(\ref{III:10}), whereas the  point  $\tau = \infty$  is  the irregular singular
point.
At present, analytical  solutions to such differential equations are known only
when the number of regular singular points does not exceed two \cite{Slavyanov}.
In our case, this means that analytical fermionic wave functions can  be  found
only for the winding  numbers  $n = \pm 1$, which correspond to the  elementary
soliton  ($n = 1$) or antisoliton ($n = -1$) of model (\ref{II:1}).

We will therefore consider the case where $n = \pm 1$.
As $\tau \rightarrow 0$,  the  radial  wave  function  $f\left(\tau\right) \sim
\tau^{l/2}$, where $l = \left\vert m - 1/2\right\vert$.
By the substitution $f\left(\tau \right)=\tau^{l/2}\left(1-\tau \right) ^{-n/2}
F\left(\tau \right)$,  differential  equation  (\ref{III:10}) is reduced to the
confluent Heun differential equation \cite{Slavyanov, Ronveaux, DLMF}
\begin{equation}
F^{\prime \prime }\left( \tau \right) +\left[ \frac{\gamma }{\tau }+\frac{
\delta }{\tau -1}+\epsilon \right] F^{\prime }\left( \tau \right) +\frac{
\alpha \tau-q}{\tau\left(\tau-1\right)}F\left( \tau \right) = 0, \label{III:12}
\end{equation}
where the parameters are
\begin{subequations}                                             \label{III:13}
\begin{eqnarray}
\alpha  &=&-\frac{1}{4}k^{2}\lambda ^{2}, \\
\gamma  &=&l+1, \\
\delta  &=&-n, \\
\epsilon  &=&0, \\
q &=&-\frac{1}{4}k^{2}\lambda ^{2}+\frac{n}{2}\left( l-\left( m-\frac{1}{2}
\right) \right).
\end{eqnarray}
\end{subequations}
Eq.~(\ref{III:12}) has two independent local solutions  in  the neighborhood of
the point $\tau = 0$: the  first  is  regular, while the other is irregular and
diverges  as $\tau^{-l}$ for $l>0$ or as $\ln(\tau)$ for $l = 0$.
To obtain the regular radial wave function, we must choose the regular solution
\begin{equation}
F\left( \tau \right) = H_{C}\left[ q,\alpha ,\gamma ,\delta ,\epsilon
,\tau \right],                                                   \label{III:14}
\end{equation}
which is called the  confluent  Heun function \cite{Slavyanov, Ronveaux, DLMF}.
In the same way, we can find  a  solution  for  the  other radial wave function
in the form  $g\left( \tau \right) = \tau^{l^{\prime }/2}\left(1 - \tau \right)
^{n/2}G\left(\tau \right)$, where $l^{\prime} = \left\vert m + 1/2\right\vert$.
The function $G(\tau)$  is  also  expressed  in  terms  of  the  confluent Heun
function
\begin{equation}
G\left( \tau \right) = H_{C} \left[q^{\prime}, \alpha^{\prime
},\gamma^{\prime},\delta^{\prime},\epsilon^{\prime},\tau\right], \label{III:15}
\end{equation}
where the parameters are
\begin{subequations}                                             \label{III:16}
\begin{eqnarray}
\alpha ^{\prime } &=&\alpha =-\frac{1}{4}k^{2}\lambda ^{2}, \\
\gamma ^{\prime } &=&l^{\prime }+1, \\
\delta ^{\prime } &=&n, \\
\epsilon ^{\prime } &=&\epsilon =0, \\
q^{\prime } &=&-\frac{1}{4}k^{2}\lambda ^{2}-\frac{n}{2}\left( l^{\prime
}+\left( m+\frac{1}{2}\right) \right).
\end{eqnarray}
\end{subequations}
The confluent  Heun  function  satisfies  the  condition $H_{C}\left[q, \alpha,
\gamma,\delta,\epsilon, 0\right] = 1$.
In the region  $\left\vert \tau \right\vert < 1$,  it  can  be  expanded into a
uniformly convergent series.
Furthermore, it can be analytically extended to the entire complex plane with a
branch cut running from $1$ to $\infty$.

Solutions  (\ref{III:14})  and  (\ref{III:15})  are  defined  up  to  arbitrary
multipliers  $\kappa_{1}$  and  $\kappa_{2}$,  respectively,  and  their  ratio
$\kappa = \kappa_{2}/\kappa_{1}$ can be  determined using differential relation
(\ref{III:9})  and  the  series  expansion  \cite{Slavyanov, Ronveaux}  of  the
confluent Heun function at the origin, as follows:
\begin{equation}
\kappa =\left\{
\begin{array}{c}
-\dfrac{\lambda }{2}\dfrac{\varepsilon - M}{m+1/2},\;m>0 \\
\dfrac{2}{\lambda }\dfrac{1/2 - m}{\varepsilon+M},\; m<0
\end{array}
\right.                                                          \label{III:17}
\end{equation}
Using the results obtained, we can write an analytical expression for the total
fermionic wave function in terms of the radial variable $\rho$:
\begin{widetext}
\begin{equation}
\psi_{m} = \mathcal{N}\left(
\begin{array}{c}
\kappa^{-1/2}
\left( \rho /\lambda \right) ^{l}\left( 1+\left( \rho /\lambda \right)
^{2}\right) ^{-n/2}H_{C}\left[ q,\alpha ,\gamma ,\delta ,\epsilon
,-\rho ^{2}/\lambda ^{2}\right] e^{i\left( m-1/2\right) \theta } \\
\kappa^{1/2} \left( \rho /\lambda \right) ^{l^{\prime }}\left( 1+\left( \rho
/\lambda \right) ^{2}\right) ^{n/2} H_{C}\left[ q^{\prime },\alpha
^{\prime },\gamma ^{\prime },\delta ^{\prime },\epsilon ^{\prime },-\rho
^{2}/\lambda ^{2}\right] e^{i\left( m+1/2\right) \theta }
\end{array}
\right) e^{-i\varepsilon t},                                     \label{III:18}
\end{equation}
\end{widetext}
where $\mathcal{N}$ is a normalization factor and the winding number $n$ of the
$\mathbb{CP}^{N-1}$ soliton can take the values $\pm 1$.

We now find  the  symmetry  properties  of  wave  function  (\ref{III:18}) with
respect to discrete transformations (\ref{III:3a}) -- (\ref{III:3c}).
It is easy to  see  that  $\psi_{\varepsilon m n}$  is  an eigenfunction of the
operators $P$ and $\Pi_{2} T$:
\begin{equation}
\left[ \psi _{\varepsilon m n}\left(t, \mathbf{x}\right) \right] ^{P}
= \left(-1\right)^{m-1/2}\psi_{\varepsilon mn}\left(t,\mathbf{x}\right),
                                                                \label{III:19a}
\end{equation}
\begin{equation}
\left[\psi_{\varepsilon mn}\left(t,\mathbf{x}\right)\right] ^{\Pi_{2}T}
 = \left( -1\right) ^{\left( 1+m/\left\vert m\right\vert \right) /2}
\psi_{\varepsilon mn}\left( t, \mathbf{x}\right),               \label{III:19b}
\end{equation}
where the eigenvalues of  the operators $H$  and $J_{3}$ and the winding number
of the $\mathbb{CP}^{N-1}$ soliton are indicated.
At the same time, the action of  the charge conjugation operator $C$ transforms
the  wave function  $\psi_{\varepsilon  m  n}$ into the wave  function  $\psi_{
-\varepsilon -m -n}$ corresponding to a negative energy state with the opposite
values  of the quantum numbers $m$ and $n$:
\begin{equation}
\left[ \psi _{\varepsilon mn}\left( t,\mathbf{x}\right) \right] ^{C}=\psi
_{-\varepsilon -m-n}\left( t,\mathbf{x}\right),                  \label{III:20}
\end{equation}
where
\begin{equation}
\psi_{-\varepsilon -m-n}\left( t,\mathbf{x}\right) =\left(
\begin{array}{c}
e^{i\left( -m-1/2\right) \theta }g\left( \rho \right)  \\
e^{i\left( -m+1/2\right) \theta }f\left( \rho \right)
\end{array}%
\right) e^{i\varepsilon t}.                                     \label{III:20a}
\end{equation}
Note that the permutation  of  the radial wave functions in Eq.~(\ref{III:20a})
compared to Eq.~(\ref{III:5}) is equivalent  to  the  replacements $\varepsilon
\rightarrow  -\varepsilon,\, m \rightarrow -m,\, n \rightarrow -n$,  as follows
from Eqs.~(\ref{III:13}) -- (\ref{III:18}).
Eq.~(\ref{III:20}) tells us that in the study of fermion-soliton systems, it is
sufficient to restrict ourselves to  fermionic ($\propto e^{-i \varepsilon t}$)
solutions,  since  antifermionic  ($\propto e^{i \varepsilon t}$) solutions are
obtained from the fermionic ones via charge conjugation.

\subsection{\label{subsec:IIIB}  Existence of fermionic bound states}

Consider the question  of  the  existence  of  fermionic  bound  states  in the
background field  of  a  $\mathbb{CP}^{N - 1}$  soliton  with winding number $n
= \pm 1$.
It is convenient to perform the substitution $f\left(\rho\right)= tilde{\rho}^{
-1/2}u\left(\tilde{\rho} \right), \,g\left( \rho \right) = \tilde{\rho} ^{-1/2}
v\left( \tilde{\rho} \right)$ to give differential equations for the new radial
functions as follows:
\begin{equation}
u^{\prime \prime}(\tilde{\rho}) - \left[\tilde{\varkappa}^{2} +
U\left(\tilde{\rho},m,n\right)\right] u(\tilde{\rho})
= 0,                                                            \label{III:21a}
\end{equation}
\begin{equation}
v^{\prime \prime}(\tilde{\rho}) - \left[\tilde{\varkappa}^{2} +
V\left(\tilde{\rho},m,n\right)\right] v(\tilde{\rho})
= 0,                                                            \label{III:21b}
\end{equation}
where $\tilde{\rho} = \rho/\lambda$, $\tilde{\varkappa}^{2} = \lambda^{2}\left(
M^{2}-\varepsilon^{2}\right)$, and the potentials
\begin{widetext}
\begin{equation}
U\left(\tilde{\rho},m,n\right) = \frac{m\left( m-1\right)}{
\tilde{\rho }^{2}}+\frac{n\left( n-2m+1\right) }{1+\tilde{\rho}^{2}}
-\frac{n\left(2+n\right)}{\left(1+\tilde{\rho}^{2}\right)^{2}}, \label{III:22a}
\end{equation}
\begin{equation}
V\left( \tilde{\rho },m,n\right) = \frac{m\left( m+1\right) }{
\tilde{\rho }^{2}}+\frac{n\left( n-2m-1\right) }{1+\tilde{\rho }^{2}}
+\frac{n\left(2-n\right)}{\left(1+\tilde{\rho}^{2}\right)^{2}}  \label{III:22b}
\end{equation}
\end{widetext}
do not depend on the scale parameter $\lambda$.
Eqs.~(\ref{III:21a}) and (\ref{III:21b}) have  the  form  of  a one-dimensional
Schr\"{o}dinger equation  with  potentials (\ref{III:22a}) and (\ref{III:22b}),
respectively.
The quantity $-\tilde{\varkappa}^{2}$  plays  the  role  of  energy and must be
negative for fermionic bound states. 
At the  same  time, the  potentials  $U$  and  $V$  have  second-order poles at
$\tilde{\rho} = 0$, and tend to zero as $\tilde{\rho} \rightarrow \infty$.
From the general properties \cite{LandauIII}  of  the Schr\"{o}dinger equation,
it follows that  for  bound  states  to  exist, $U$  and $V$ must take negative
values.
An analysis shows that both $U$ and $V$ have areas of negative  values only for
$ m = 1/2,\, n = 1$ and $m = -1/2,\, n = -1$.
For other values of $m$ and $n$, at  least  one  of $U$ and $V$ turns out to be
positive for all  $\tilde{\rho} \in (0, \infty)$,  which  makes  the  existence
of bound fermionic states impossible.

Consider one of the possible cases, say $m = 1/2,\, n = 1$.
Another possible case, $m=-1/2,\, n=-1$, is reduced to the previous one through
the relation $U\left(\tilde{\rho},m,n\right)=V\left(\tilde{\rho},-m,-n\right)$.
It follows from Eq.~(\ref{III:22a})  that  the  potential $U\left(\tilde{\rho},
1/2, 1 \right) = -\left(2\tilde{\rho}\right)^{-2} - 3\left(1 + \tilde{\rho}^{2}
\right)^{-2}+\left(1+\tilde{\rho}^{2}\right)^{-1}$.
Furthermore,  Eq.~(\ref{III:21a}) admits a mechanical analogy; it describes the
one-dimensional motion  of  a  unit  mass  particle  along the $u$-axis in time
$\tilde{\rho}$.
The motion of the particle occurs under the action of the time-dependent linear
force $F(\tilde{\rho})=\left(\tilde{\varkappa}^{2} + U\left(\tilde{\rho},1/2, 1
\right)\right)u(\tilde{\rho})$.
Since for $m = 1/2$  the  solution  $u(\tilde{\rho}) \sim\tilde{\rho}^{1/2}$ as
$\tilde{\rho} \rightarrow 0$, the  particle  has  the  coordinate  $u  = 0$ and
possesses an infinite speed at the initial time $\tilde{\rho} = 0$.
This infinite  speed,  however, is  compensated  by  the  action  of  the force
$F(\tilde{\rho})$, which also tends to infinity as $\tilde{\rho}\rightarrow 0$.

We now consider the limiting case $\tilde{\varkappa}^{2}=0$,  which corresponds
to $\varepsilon = \pm M$.
It is easy to see that in this  case,  the  system  of first-order differential
equations in  Eqs.~(\ref{III:6a}) and (\ref{III:6b}) splits,  and its solutions
can therefore be expressed in terms of the elementary functions
\begin{subequations}                                             \label{III:23}
\begin{eqnarray}
\psi _{M\frac{1}{2}1} &=&
\begin{pmatrix}
\left( \lambda ^{2}+\rho ^{2}\right) ^{-1/2} \\
0%
\end{pmatrix}%
e^{-iMt},                                                       \label{III:23a}
 \\
\psi _{-M-\frac{1}{2}-1} &=&
\begin{pmatrix}
0 \\
\left( \lambda ^{2}+\rho ^{2}\right) ^{-1/2}
\end{pmatrix}
e^{iMt}.                                                        \label{III:23b}
\end{eqnarray}
\end{subequations}
Eqs.~(\ref{III:23a})    and    (\ref{III:23b})    are     special    cases   of
Eq.~(\ref{III:18}).
It follows from Eq.~(\ref{III:17}) that  the  multiplier $\kappa$ tends to zero
(infinity) when $\varepsilon \rightarrow  M$  and  $m = 1/2$ ($\varepsilon
\rightarrow -M$ and $m = -1/2$).
The infinity arising in the upper (lower) component of fermionic wave  function
(\ref{III:18}) is compensated,  since the normalization factor $\mathcal{N}$ is
proportional to $\kappa^{1/2}$ ($\kappa^{-1/2}$).
As  a  result,  the  lower  (upper)   component   of  fermionic  wave  function
(\ref{III:18})    vanishes,    and     we    arrive   at    Eq.~(\ref{III:23a})
(Eq.~(\ref{III:23b})).
In this case, the  confluent  Heun  function  that  corresponds  to the nonzero
component of fermionic wave  function (\ref{III:18}) degenerates to a constant.

It  follows  from   Eqs.~(\ref{III:23a})  and  (\ref{III:23b})  that  at  large
distances from the soliton, the solutions $\psi_{\pm M\,\pm 1/2\,\pm 1} \propto
\rho^{-1}$.
Hence, the solutions $\psi_{\pm M\, \pm 1/2\, \pm 1}$ cannot be normalized, and
therefore  cannot  be  regarded  as  a  part  of  the discrete  spectrum of the
Hamiltonian (\ref{III:2}).
From Eq.~(\ref{III:23a}), we obtain the solution
\begin{equation}
u_{M\frac{1}{2}1}\left(\tilde{\rho}\right) = \tilde{\rho}^{1/2}
\left(1+\tilde{\rho}^{2}\right)^{-1/2}                           \label{III:23}
\end{equation}
to Eq.~(\ref{III:21a}).
It follows from Eq.~(\ref{III:23}) that  the  solution $u_{M\frac{1}{2}1}\left(
\tilde{\rho} \right)$  increases monotonically  from  zero to $2^{-1/2}$ on the
interval $(0, 1)$ and then decreases monotonically to zero on the interval $(1,
\infty)$.
Note that the solution $u_{M\frac{1}{2}1}\left(\tilde{\rho}\right) \sim \tilde{
\rho}^{-1/2}$ as  $\tilde{\rho} \rightarrow \infty$.

Next, we  define  the  effective  potential $U_{\text{eff}}\left( \tilde{\rho},
\tilde{\varkappa} \right) = \tilde{\varkappa}^{2} + U\left(\tilde{\rho}, 1/2, 1
\right)$, where $\tilde{\varkappa}^{2} = \lambda^{2}\left( M^{2} -\varepsilon^{
2}\right)$ must be positive for fermionic bound states.
The effective  potential  $U_{\text{eff}}\left( \tilde{\rho}, \tilde{\varkappa}
\right)$  increases  monotonically  from  $-\infty$  to  $0.0428454  +  \tilde{
\varkappa}^{2}$   on   the   interval   $\left(0,   2.79921\right)$,  and  then
decreases  monotonically to  $\tilde{\varkappa}^{2}$  on  the  interval $\left(
2.79921, \infty \right)$.
In  addition,  $U_{\text{eff}} \left( \tilde{\rho}, \tilde{ \varkappa} \right)$
vanishes  at  $\tilde{\rho}   =   \tilde{\rho}_{0}( \tilde{\varkappa} )$, where
$\tilde{\rho}_{0}(0) \approx 1.85216$ and $\tilde{\rho}_{0}(\tilde{\varkappa})$
decreases monotonically with an increase in $\tilde{\varkappa}$.
It follows that the force $F(\tilde{\rho}) = U_{\text{eff}}\left( \tilde{\rho},
\tilde{\varkappa} \right)u(\tilde{\rho})$  is  attractive when $\tilde{\rho}\in
\left(0,\tilde{\rho}_{0}( \tilde{\varkappa})\right)$,  and  is  repulsive  when
$\rho\in\left(\tilde{\rho}_{0}(\tilde{\varkappa}), \infty \right)$.
This means  that  in  order   to   correspond   to  a  ground  state  of energy
$\varepsilon_{0}$, the trajectory $u_{\varepsilon_{0}\frac{1}{2} 1}(\tilde{\rho
})$ of the particle must reach a maximum at some point $\tilde{\rho}_{\text{max
}}<\tilde{\rho}_{0}(\tilde{\varkappa}_{0})$  in  the  region  of attraction and
then  decrease monotonically,  tending  to  zero  as  $\tilde{\rho} \rightarrow
\infty$.
The monotonic decrease of  $u_{\varepsilon_{0} \frac{1}{2} 1}(\tilde{\rho})$ is
due to the fact that the radial wave function of the ground state has no nodes.
In addition, Eq.~(\ref{III:21a}) tells us that  $u_{\varepsilon_{0} \frac{1}{2}
1}(\tilde{\rho})  \sim  \exp(-\tilde{\varkappa}_{0}\tilde{\rho})$  as  $\tilde{
\rho}  \rightarrow  \infty$,  where  $\tilde{\varkappa}_{0}^{2}  =  \lambda^{2}
\left(M^{2}-\varepsilon_{0}^{2}\right)$.

As $\tilde{\varkappa}^{2}$ increases, the region of attraction $\left(0,\tilde{
\rho}_{0}(\tilde{\varkappa}) \right)$  of  $U_{\text{eff}} \left( \tilde{\rho},
\tilde{\varkappa}  \right)$   decreases   while   the   region   of   repulsion
$\left(\tilde{\rho}_{0}(\tilde{\varkappa}), \infty \right)$ increases.
Since the  force  $F(\tilde{\rho}) = U_{\text{eff}} \left(\tilde{\rho}, \tilde{
\varkappa}\right)u(\tilde{\rho})=\left(\tilde{\varkappa}^{2}+U\left(\tilde{\rho
},1/2, 1\right)\right) u(\tilde{\rho})$, the attraction force decreases and the
repulsion force increases with the  growth  of  $\tilde{\varkappa}^{2}$.
We can normalize  the  wave  function $u_{\varepsilon_{0}\frac{1}{2} 1}(\tilde{
\rho})$ of the assumed bound state by the condition $u_{\varepsilon_{0}\frac{1}
{2} 1}(\tilde{\rho})/u_{M\frac{1}{2}1}\left( \tilde{\rho}\right)\rightarrow  1$
as $\tilde{\rho} \rightarrow 0$.
It then follows from the above that the  trajectories  $u_{M\frac{1}{2}1}\left(
\tilde{\rho}\right)$ and $u_{\varepsilon_{0} \frac{1}{2} 1}(\tilde{\rho})$ must
satisfy the  inequality  $u_{\varepsilon_{0} \frac{1}{2}1}(\tilde{\rho}) > u_{M
\frac{1}{2} 1} (\tilde{\rho})$.
Recall, however, that $u_{M\frac{1}{2} 1}\left(\tilde{\rho}\right) \sim \tilde{
\rho}^{-1/2}$  and  $u_{\varepsilon_{0}\frac{1}{2} 1}(\tilde{\rho})  \sim \exp(
- \tilde{\varkappa}_{0}\tilde{\rho})$  as  $\tilde{\rho}  \rightarrow  \infty$,
and  hence  the  ratio   $u_{\varepsilon_{0} \frac{1}{2} 1}(\tilde{\rho})/u_{M
\frac{1}{2} 1}(\tilde{\rho})$   must   tend  to  zero  in  this  limit,  which
contradicts the condition $u_{\varepsilon_{0}\frac{1}{2}1}(\tilde{\rho})  > u_{
M \frac{1}{2} 1}(\tilde{\rho})$.
We can conclude that there are no bound fermionic  states  with quantum numbers
$m = 1/2, \, n = 1$.
It follows that there are no  bound  fermionic  states  in the background field
of a $\mathbb{CP}^{N-1}$ soliton with $n = \pm 1$.

\subsection{\label{subsec:IIIС}  General formalism for fermion scattering}

We now turn to the description of fermion  scattering  in  the background field
of a $\mathbb{CP}^{N-1}$ soliton.
For a fermion with  initial  momentum  $\mathbf{k} = (k, 0)$,  according to the
general principles of  the  theory  of scattering \cite{LandauIII, Taylor}, the
asymptotics of the wave function of the fermionic scattering state has the form
\begin{equation}
\Psi \sim \psi _{\varepsilon ,\mathbf{k}\,}+\frac{1}{\sqrt{2\varepsilon }}
u_{\varepsilon ,\mathbf{k}^{\prime }}f\left( k,\theta \right) \frac{
e^{ik\rho }}{\sqrt{-i\rho }},                                    \label{III:24}
\end{equation}
where
\begin{equation}
\psi _{\varepsilon ,\mathbf{k}}=\frac{1}{\sqrt{2\varepsilon}}
\begin{pmatrix}
\sqrt{\varepsilon + M} \\
i\sqrt{\varepsilon - M}
\end{pmatrix}
e^{-i k x}                                                       \label{III:25}
\end{equation}
is the wave function of the incoming fermion with momentum $\mathbf{k}=(k, 0)$,
\begin{equation}
u_{\varepsilon ,\mathbf{k}^{\prime }}=
\begin{pmatrix}
\sqrt{\varepsilon + M} \\
i\sqrt{\varepsilon - M}e^{i\theta }                              \label{III:26}
\end{pmatrix}
\end{equation}
is the spinor amplitude of  the  wave  function  of  the  outgoing fermion with
momentum $\mathbf{k}^{\prime} = (k\cos(\theta), k\sin(\theta))$, and $f\left(k,
\theta \right)$ is the scattering amplitude.

The scattering amplitude $f\left(k, \theta \right)$ can be expanded in terms of
the partial scattering amplitudes $f_{m}\left( k\right)$ as
\begin{equation}
f\left( k,\theta \right) =\sum\limits_{m}f_{m}\left( k\right) e^{i\left(
m-1/2\right)\theta},                                             \label{III:27}
\end{equation}
where the summation is  taken  over  the  half-integer  eigenvalues  of angular
momentum (\ref{III:4}).
The partial scattering  amplitudes  can  in turn be  written  in  terms  of the
partial elements of the $S$-matrix as
\begin{equation}
f_{m}\left( k\right) =\frac{1}{i\sqrt{2\pi k}}\left( S_{m}\left( k\right)
-1\right).                                                       \label{III:28}
\end{equation}
Similarly, wave function (\ref{III:24}) can  also  be  decomposed  into partial
waves as $\Psi = \sum\nolimits_{m}\psi_{m}$.
The asymptotic behavior of the partial  waves can be expressed in terms of the
partial elements of the $S$-matrix as follows:
\begin{widetext}
\begin{equation}
\psi _{m} \sim \frac{\left( -1\right) ^{1/4}}{\sqrt{2\pi k\rho }}
\begin{pmatrix}
-i\sqrt{\frac{\varepsilon + M}{2\varepsilon }}\left[ i\left( -1\right)
^{m-1/2}e^{-ik\rho }+S_{m}e^{ik\rho }\right] e^{i\left( m-1/2\right)\theta}
\\
\sqrt{\frac{\varepsilon - M}{2\varepsilon }}\left[ i\left( -1\right)
^{m+1/2}e^{-ik\rho }+S_{m}e^{ik\rho }\right] e^{i\left( m+1/2\right)\theta}
\end{pmatrix}\!.                                                 \label{III:29}
\end{equation}
\end{widetext}

Using standard methods from the theory  of scattering \cite{LandauIII, Taylor},
we can write the differential cross-section  for the elastic fermion scattering
in terms of the scattering amplitude $f\left(k, \theta \right)$ as
\begin{equation}
d\sigma/d\theta =\left\vert f\left(k, \theta \right) \right\vert^{2}.
                                                                 \label{III:30}
\end{equation}
In turn, the   partial  cross-sections  for  the elastic fermion scattering are
expressed in terms of the partial scattering amplitudes as
\begin{equation}
\sigma _{m}=2\pi \left\vert f_{m}\left( k\right) \right\vert
^{2}=k^{-1}\left\vert S_{m}\left( k\right) -1\right\vert^{2}.    \label{III:31}
\end{equation}
Note that in  $(2 + 1)$  dimensions,  the  cross-sections $d\sigma/d\theta$ and
$\sigma _{m}$ have the dimension of length \cite{LandauIII}.
The unitarity of  the $S$-matrix, $S S^{\dagger} = S^{\dagger} S = \mathbb{I}$,
results in  the  unitarity  condition  for  the  partial  $S$-matrix  elements,
$\left\vert S_{m}\left( k \right) \right\vert = 1$.
This condition allows us to express  the  partial  $S$-matrix  elements $S_{m}$
in terms of the partial phase shifts $\delta_{m}$ as
\begin{equation}
S_{m}\left( k\right) = e^{2 i \delta _{m}\left( k\right)}.       \label{III:32}
\end{equation}

The scale invariance of the bosonic  sector  of model (\ref{II:1}) leads to the
existence of the parameter  $\lambda$,  which  determines the effective size of
soliton  solution   (\ref{II:6}),   and   hence   affects  the  fermion-soliton
scattering.
It follows from a dimensional analysis and Eqs.~(\ref{III:13}) -- (\ref{III:18})
that the dependence of $S_{m}$ on the momentum $k=(\varepsilon^{2}-M^{2})^{1/2}
$  and the scale parameter  $\lambda$  enters  only  through  the dimensionless
combination $\tilde{k} = k \lambda$.

The phase shifts  $\delta_{m}(\tilde{k})$  are determined only by the arguments
of the confluent Heun functions in Eq.~(\ref{III:18}), and do not depend on the
prefactors.
These  arguments  depend  on  the  parameter  $\varepsilon$  only  through  the
momentum   squared  $k^{2}  =  \varepsilon^{2}  -  M^{2}$,  i.e.,  only through
$\varepsilon^{2}$.
It follows  that  the  phase  shifts  $\delta_{m}(\tilde{k})$  are the same for
both the fermionic ($\propto e^{-i\varepsilon t}$) and  antifermionic ($\propto
e^{i \varepsilon t}$) solutions.
Further, charge conjugation (\ref{III:3a}) is  reduced  to a permutation of the
wave function components and  to their complex conjugation, which cannot change
the phase shifts.
This  is   because   the   arguments   of   the   confluent  Heun  functions in
Eq.~(\ref{III:18}) are  real,  meaning  that  these  functions  are  also real.
From this and Eq.~(\ref{III:20}),  we  come  to  the  conclusion that the phase
shifts satisfy the relation
\begin{equation}
\delta_{m n}(\tilde{k}) = \delta_{-m\, -n}(\tilde{k}),           \label{III:33}
\end{equation}
where the dependence of the phase shift on the soliton winding number $n=\pm 1$
is indicated.

\section{\label{sec:IV} Fermion scattering in the Born approximation}

In  Sec.~\ref{sec:III},  we   were   able   to   obtain  analytical  expression
(\ref{III:18}) for the fermionic wave functions in the background field  of the
$\mathbb{CP}^{N - 1}$ soliton for winding numbers $n = \pm 1$.
The  next  step  would  be  to  obtain an exact  analytical  expression for the
scattering amplitude (\ref{III:27}).
To do this, according to Eqs.~(\ref{III:28}) and (\ref{III:32}), we need to know
exact analytical expressions for the partial phase shifts $\delta_{m}(k)$.
However, unlike the well-studied Bessel functions, exact analytical expressions
$\delta_{m}(k)$ are unknown  for  the  confluent  Heun  functions  appearing in
Eq.~(\ref{III:18}).
Hence, we  cannot  obtain  an  exact  analytical  expression for the scattering
amplitude.
In view of this, it is important  to  study  the fermion scattering in the Born
approximation, which gives us  a  chance  to  obtain  an approximate analytical
expression for the scattering amplitude.

It follows from Eqs.~(\ref{II:1}) and  (\ref{II:2})  that  the  fermion-soliton
interaction is described by the potential term
\begin{equation}
V_{\text{int}}=\bar{\psi}_{a}\gamma^{\mu}A_{\mu}\psi_{a}.          \label{IV:I}
\end{equation}
In the  background  field  approximation,  the  gauge  field  $A_{\mu}$ defined
by Eq.~(\ref{II:7}) does not depend on the fermion fields $\psi_{a}$.
It follows from this and Eq.~(\ref{IV:I})  that all components of the fermionic
multiplet $\left(\psi_{1},\ldots, \psi_{N}\right)$  interact with the $\mathbb{
CP}^{N-1}$ soliton in the same way and independently of each other.

Using Eq.~(\ref{IV:I}), we  can  write  the  first-order Born amplitude for the
fermion-soliton scattering as follows:
\begin{equation}
f\left(\mathbf{k}^{\prime},\mathbf{k}\right) =
-\left(8\pi k\right)^{-1/2}\bar{u}
_{\varepsilon,\mathbf{k}^{\prime}}\mathbb{\gamma}^{\mu}A_{\mu}\left(
\mathbf{q}\right) u_{\varepsilon ,\mathbf{k}},                     \label{IV:1}
\end{equation}
where
\begin{equation}
A_{\mu }\left( \mathbf{q}\right) = \int A_{\mu}\left(\mathbf{x}\right)
e^{-i\mathbf{q \cdot x}}d^{2}x                                     \label{IV:2}
\end{equation}
and $\mathbf{q} = \mathbf{k}^{\prime} - \mathbf{k}$  is  the momentum transfer.
The Born amplitude (\ref{IV:1}) can be expressed in an analytical form.
For winding  numbers  $\left\vert n \right \vert \ge 2$,  the Born amplitude is
expressed in terms of the Meijer $G$-functions \cite{PBM}; however, for winding
numbers $n=\pm1$, corresponding to the elementary $\mathbb{CP}^{N-1}$ solitons,
the Born  amplitude can be written in terms of modified Bessel functions of the
second kind:
\begin{eqnarray}
f\left( \mathbf{k}^{\prime },\mathbf{k}\right)  &=&i n \sqrt{2\pi}
k^{1/2}\lambda\,\text{sign}\left(\vartheta_{2}-\vartheta_{1}\right) \nonumber
  \\
&&\times e^{-i\left(\vartheta_{2}-\vartheta_{1}\right)/2}
\text{K}_{1}\left(q\lambda \right),                                \label{IV:3}
\end{eqnarray}
where  the  angle  $\vartheta_{1}$   ($\vartheta_{2}$)  defines  the  direction
of  motion  of  the ``in" (``out")  fermion, $q  =  2  k  \sin\left( \left\vert
\vartheta_{2}  -  \vartheta_{1} \right\vert/2 \right)$  is the magnitude of the
momentum transfer, and $n = \pm 1$ is the winding number of the $\mathbb{CP}^{N
- 1}$ soliton.
The amplitude of antifermion  scattering differs only in terms of its sign from
the amplitude of fermion scattering in Eq.~(\ref{IV:3}).
Using known  criteria \cite{LandauIII,Taylor},  it  can  be shown that the Born
approximation  is  applicable  under  the  following conditions
\begin{equation}
k\lambda \gg 1\;\text{and\ }
\left\vert \vartheta_{2} - \vartheta_{1} \right\vert \ll
\left( k \lambda \right)^{-1/2} \ll 1.                            \label{IV:3a}
\end{equation}
It follows from Eq.~(\ref{IV:3a}) that the  Born  approximation is suitable for
describing the low-angle scattering of high-energy fermions.

Eq.~(\ref{IV:3}) tells  us  that  the  amplitude  $f\left( \mathbf{k}^{\prime},
\mathbf{k}\right)$ is  Hermitian with respect to the permutation of the fermion
momenta
\begin{equation}
f\left(\mathbf{k}^{\prime},\mathbf{k}\right) = f^{\ast}\left(\mathbf{k},
\mathbf{k}^{\prime}\right),                                        \label{IV:4}
\end{equation}
as it should be in the Born approximation \cite{LandauIII, Taylor}.
Another symmetry relation
\begin{equation}
f\left( \mathbf{k}^{\prime },\mathbf{k}\right) =
f( \tilde{\mathbf{k}},\tilde{\mathbf{k}}^{\prime }),               \label{IV:5}
\end{equation}
where $\tilde{\mathbf{k}}=\left(k_{x},-k_{y}\right)$ and $ \tilde{\mathbf{k}}^{
\prime} = \left( k_{x}^{\prime },  -k_{y}^{\prime} \right)$,  follows  from the
invariance of the Dirac equation (\ref{III:1}) under  $\Pi_{2}T$ transformation
(\ref{III:3c}).
As already mentioned, the  scattering  amplitude  for  antifermions is obtained
from Eq.~(\ref{IV:3}) by the replacement $n \rightarrow -n$.
It follows that the scattering  of  an  antifermion  on the $\mathbb{CP}^{N-1}$
soliton with winding number  $n = \pm 1$  is  equivalent  to  the scattering of
a fermion on the $\mathbb{CP}^{N-1}$  soliton  with  winding  number $n=\mp 1$,
which corresponds to Eq.~(\ref{III:20}).

Using Eq.~(\ref{IV:3}) and  the  known  asymptotic forms of the modified Bessel
function $\text{K}_{1}\left( q  \lambda \right)$,  we  can  study  the behavior
of the Born amplitude  for large and small values of the momentum transfer $q$.
For large momentum transfers, we find that the Born amplitude
\begin{eqnarray}
f &\sim & i n \pi \left( \lambda /2\right) ^{1/2}\text{sign}\left( \vartheta
_{2}-\vartheta_{1}\right)e^{-i\left(\vartheta_{2}-\vartheta_{1}\right)
/2}                                                                 \nonumber
 \\
&&\times e^{-\lambda q}\sin \left( \left\vert \vartheta _{2}-\vartheta
_{1}\right\vert /2\right) ^{-1/2},                                 \label{IV:6}
\end{eqnarray}
where the angles $\vartheta_{1}$ and $\vartheta_{2}$ are fixed and $\vartheta_{
1} \ne \vartheta_{2}$.
It  follows  from   Eq.~(\ref{IV:6})   that   the   Born   amplitude  decreases
exponentially with  an  increase  in the dimensionless combination $\lambda q$.
We now consider the  case  of  low momentum transfer $q$ and high fixed fermion
momentum $k$, which  corresponds  to small scattering  angles $\Delta \vartheta \equiv\left\vert\vartheta_{2}-\vartheta_{1}\right\vert=2\arcsin\left[q/\left(2k
\right) \right] \approx q/k$.
In this case, the asymptotics of the Born amplitude is
\begin{equation}
f \sim i n \sqrt{2\pi }k^{1/2}q^{-1} + n\sqrt{\pi /2}k^{-1/2}.     \label{IV:7}
\end{equation}
We see that in the limit of small $q$, the Born amplitude diverges as $q^{-1}$.
Furthermore, unlike Eq.~(\ref{IV:6}),  the  leading  asymptotic  terms shown in
Eq.~(\ref{IV:7}) do  not  depend  on  the  parameter $\lambda$  determining the
soliton size.

Next, we turn to the study  of  the  partial amplitudes $f_{m}\left( k \right)=
\left(2\pi\right)^{-1}\int\nolimits_{0}^{2\pi}e^{-i\left(m-1/2\right)\vartheta}
f\left( k,\vartheta \right) d\vartheta$  corresponding  to  the  Born amplitude
(\ref{IV:3}).
The imaginary part of the integrand diverges as $\vartheta\rightarrow 0,\,2\pi$
and is odd  with  respect  to  $\vartheta = \pi$,  and  hence the corresponding
integral vanishes in the sense of the principal value.
The real part of the integrand is  finite and even with respect to $\vartheta =
\pi$, meaning that the corresponding integral exists and is nonzero.
From this result, it is easy to show that  the  partial Born amplitudes are odd
under the replacement $m \rightarrow -m$, i.e.,
\begin{equation}
f_{m\,}\left( k\right) = -f_{-m\,}\left( k\right).                 \label{IV:8}
\end{equation}

The partial Born amplitudes $f_{m\,}\left( k \right)$ can be expressed in terms
of the Meijer $G$-functions \cite{PBM}.
These expressions, however,  can  be  significantly  simplified in the limit of
large fermion momenta $k$ as
\begin{equation}
f_{m}\sim n m \sqrt{\pi /2}\lambda ^{-1}k^{-3/2}.                  \label{IV:9}
\end{equation}
From Eqs.~(\ref{III:28}) and (\ref{IV:9}), we  can  obtain asymptotic forms for
the partial $S$-matrix elements and  phase  shifts  in  the Born approximation:
\begin{equation}
S_{m} = e^{2 i \delta_{m}} \sim 1 + i\frac{n m\pi}{ k \lambda}    \label{IV:10}
\end{equation}
and
\begin{equation}
\delta_{m} \sim  \frac{n m \pi}{2 k \lambda}.                     \label{IV:11}
\end{equation}
It follows from Eq.~(\ref{IV:11}) that  the  phase  shifts $\delta_{m}$ tend to
zero as $k \rightarrow\infty$, which is consistent with the basic principles of
scattering theory \cite{LandauIII, Taylor}.

Using  Eq.~(\ref{IV:3}),  we  obtain   an    expression  for  the  differential
cross-section of the fermion scattering in the Born approximation
\begin{equation}
d\sigma /d\vartheta =2\pi k\lambda^{2}\text{K}_{1}\left( 2k\lambda \sin
\left( \vartheta /2\right) \right)^{2}.                           \label{IV:13}
\end{equation}
We see that $d\sigma /d\vartheta \sim 2\pi k^{-1}\vartheta ^{-2}$ as $\vartheta
\rightarrow 0$.
At the same time, $d\sigma /d\vartheta \sim 2\pi k^{-1}(2\pi - \vartheta)^{-2}$
as $\vartheta \rightarrow 2 \pi$.
It follows  that  the  total  cross-section $\sigma = \int \nolimits_{0}^{2\pi}
\left( d\sigma   /  d\vartheta  \right)  d\vartheta$   of  the  fermion-soliton
scattering diverges at the lower and upper limits of the integral.
However, the  transport cross-section $\sigma _{\text{tr}}=\int\nolimits_{0}^{2
\pi }\left( 1-\cos \left( \vartheta \right) \right)  \left( d\sigma /d\vartheta
\right) d\vartheta$ is finite, and  can  be  expressed  in terms of the  Meijer
$G$-functions,  defined  according  to Ref.~\cite{PBM}, as
\begin{equation}
\sigma _{\text{tr}}=4\pi ^{2}k\lambda ^{2}G_{2,4}^{3,1}\left( 4k^{2}\lambda
^{2}\left\vert
\begin{array}{c}
-\frac{1}{2},\frac{1}{2} \\
-1,0,1,-1%
\end{array}%
\right. \right).                                                  \label{IV:14}
\end{equation}
Using known asymptotic expansion  for  the  Meijer  $G$-function, we obtain the
asymptotics of the transport cross-section  for  large  fermion momentum $k$ as
\begin{equation}
\sigma_{\text{tr}}\sim \frac{3\pi^{3}}{16k^{2}\lambda} +
O\left(\lambda^{-3}k^{-4}\right).                                 \label{IV:15}
\end{equation}

\section{\label{sec:V} Numerical results}

In Sec.~\ref{sec:III}, we  found  an  exact  solution  for  the  fermionic wave
function in the background  field  of  the  $\mathbb{CP}^{N - 1}$  soliton with
winding number $n = \pm 1$.
This exact solution is  expressed  in  terms  of  the  confluent Heun functions
\cite{Slavyanov, Ronveaux, DLMF}.
We now want to find  the  partial  phase  shifts  $\delta_{m}(\tilde{k})$ for a
range of values for the dimensionless  combination  $\tilde{k} = k \lambda$, as
these will give the most complete description of the fermion scattering.
Since there is no analytic  form  for  the  asymptotics  of  the confluent Heun
function in the region of large $\rho$,  we  need  to  use numerical methods to
solve this problem.

Exact solution (\ref{III:18})  and  general  asymptotic form (\ref{III:29}) are
two-component spinors.
Let us define the ratio of the spinor components taken at two successive points
$\rho$ and $\rho + \Delta \rho$ as
\begin{equation}
r_{\varepsilon mn}^{i}\left(\rho ,\Delta \rho \right) =
\psi_{\varepsilon mn}^{i}\left(\rho+\Delta\rho\right)/\psi_{\varepsilon mn}^{i}
\left( \rho \right),                                                \label{V:1}
\end{equation}
where the index $i = 1, 2$ numbers the spinor components.
For sufficiently large $\rho$,  exact  solution (\ref{III:18}) tends to general
asymptotic form (\ref{III:29}).
It follows that  in  this case, the ratio $r_{\varepsilon m n}^{i}$  calculated
with   Eq.~(\ref{III:18})   must    be    close   to   that   calculated   with
Eq.~(\ref{III:29}).
Equating these two ratios calculated for some $\rho  \gg  \lambda$  and $\Delta
\rho \sim \lambda$, we obtain an  approximate equation to determine the partial
$S$-matrix element $S_{m} = \exp(2 i \delta_{m})$ in Eq.~(\ref{III:29}).
Since we can use  both $r_{\varepsilon m n}^{1}$  and $r_{\varepsilon m n}^{2}$
for this purpose, we have two approximate equations determining $S_{m}$.
When  $\rho \gg \lambda$, $k \rho \gg 1$,  and  $\Delta \rho \sim \lambda$, the
solutions to these two equations  become  very  close to each other, and tend to
the same limit as $\rho \rightarrow \infty$.
We used the arithmetic mean  of  these  two  solutions as a numerical value for
$S_{m}$.
To calculate the confluent Heun functions  for large values of their arguments,
the  highly   efficient   numerical   algorithms   of  the   {\sc{Mathematica}}
\cite{Mathematica}  and  {\sc{Maple}} \cite{Maple} software packages were used.

\begin{figure}[tbp]
\includegraphics[width=0.5\textwidth]{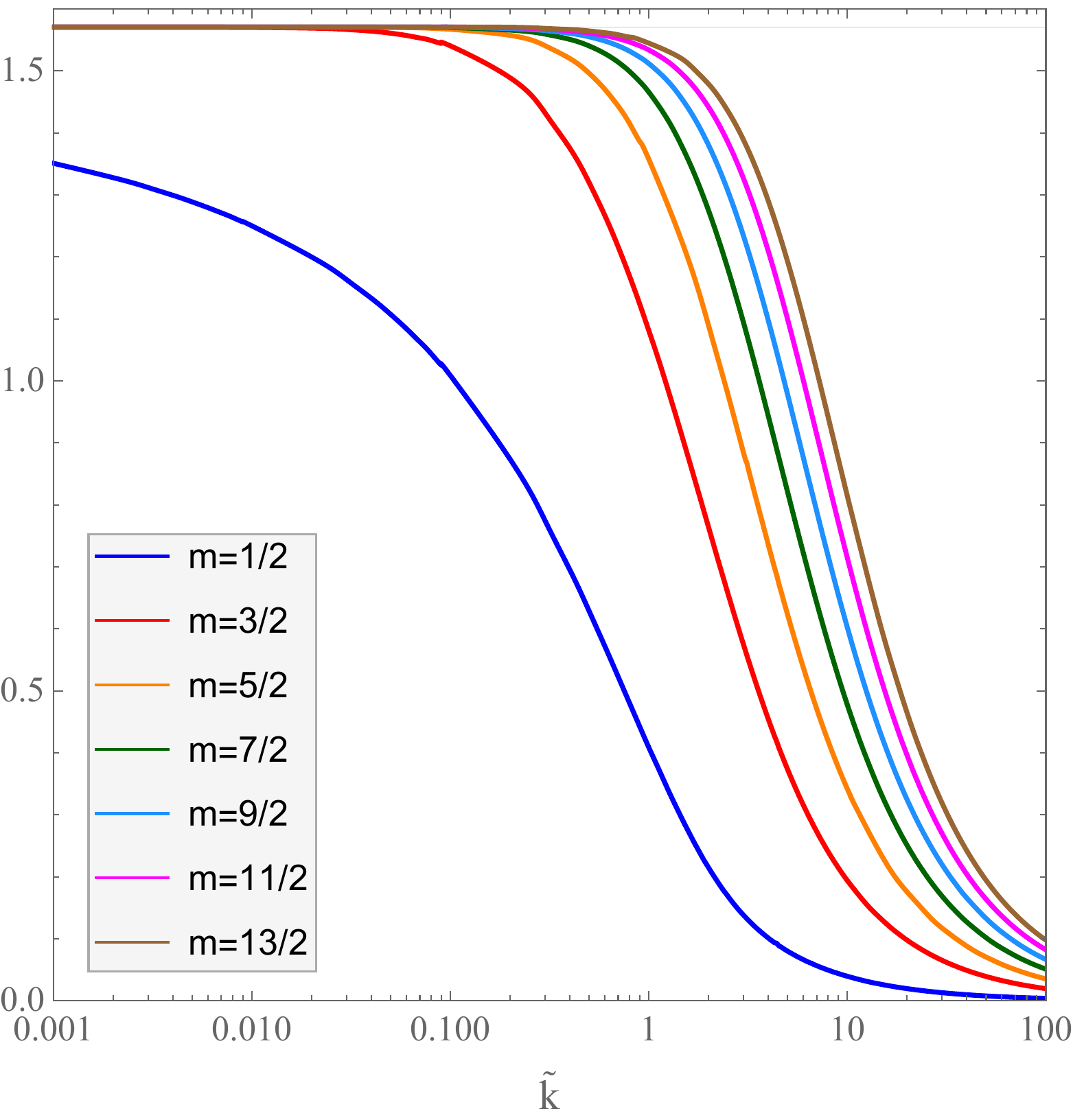}
\caption{\label{fig1}    Dependence of  the  phase shifts $\delta_{m n}$ on the
dimensionless combination $\tilde{k}=k\lambda$ for $m = 1/2,\,3/2,\, 5/2,\,7/2,
\,9/2,\,11/2,\,13/2$, and $n = 1$.}
\end{figure}

\begin{figure}[tbp]
\includegraphics[width=0.5\textwidth]{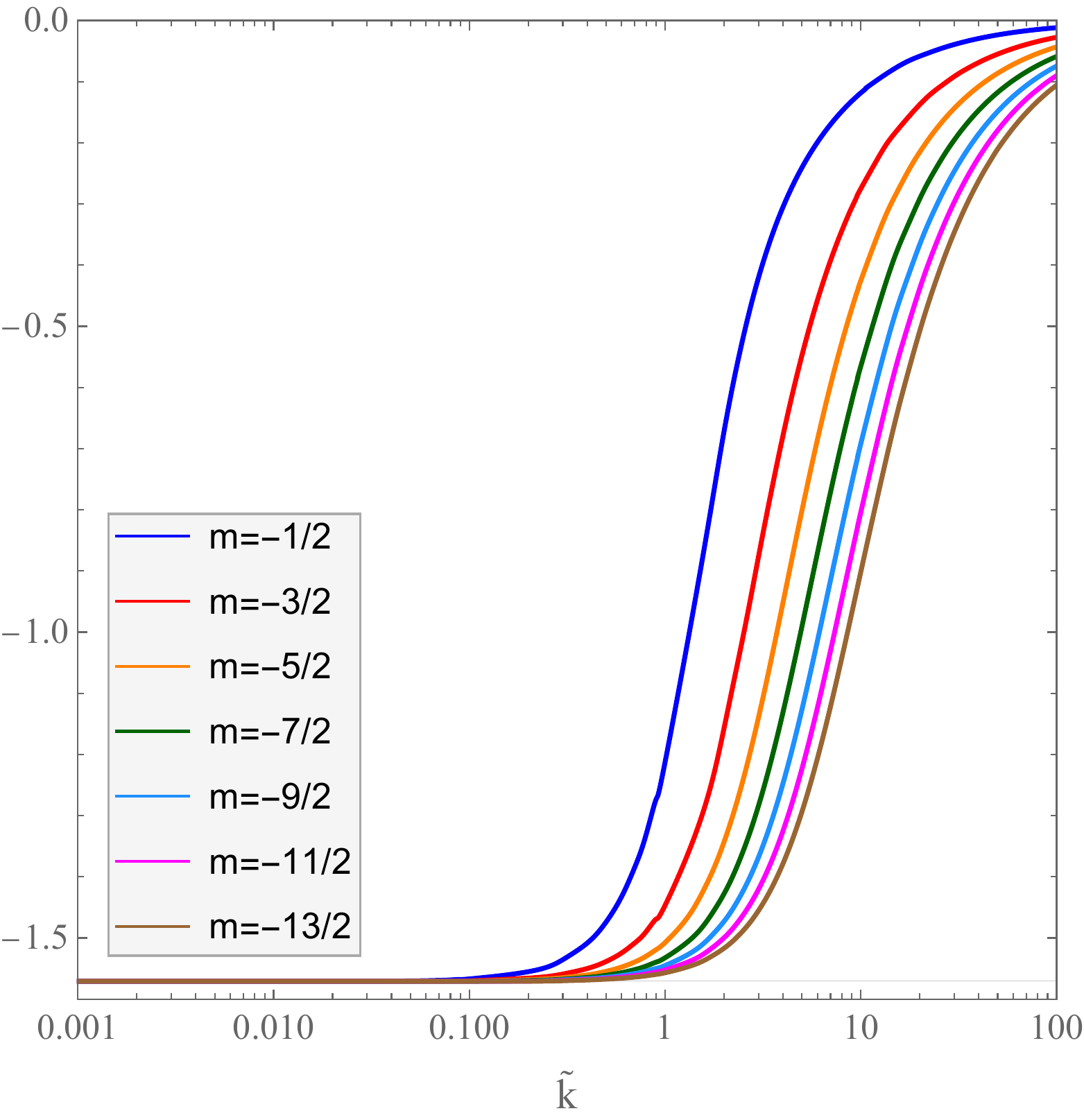}
\caption{\label{fig2}    Dependence of  the  phase shifts $\delta_{m n}$ on the
dimensionless combination $\tilde{k} = k\lambda$ for $m = -1/2,\,-3/2,\,-5/2,\,
-7/2,\,-9/2,\,-11/2,\,-13/2$, and $n = 1$.}
\end{figure}

Figure~\ref{fig1} shows the dependences  of  the phase shifts $\delta_{m n}$ on
the dimensionless  combination  $\tilde{k} = k\lambda$ for the angular momentum
eigenvalues $m = 1/2,\,3/2,\,5/2,\,7/2,\, 9/2, \,11/2, \,13/2$, and the soliton
winding number $n = 1$.
Similarly, Fig.~\ref{fig2} shows  the curves $\delta_{m n}(\tilde{k})$ for $m =
-1/2,\,-3/2,\,-5/2,\,-7/2,\,-9/2,\,-11/2,\,-13/2$, and $n = 1$.
Eq.~(\ref{III:33})  tells   us  that  the  curves $\delta_{m\,- 1}(\tilde{k}) =
\delta_{-m\, 1}(\tilde{k})$, and  these  can  therefore  be  obtained  from the
curves shown in Figs.~\ref{fig1} and \ref{fig2}.
We checked Eq.~(\ref{III:33}) using numerical methods.
It follows  from  Eq.~(\ref{III:33})  and  Figs.~\ref{fig1} and \ref{fig2} that
$\delta_{m n}\left(0\right) = \text{sign}\left( m n\right) \pi/2$.
Note that in  the  case  of short-range  forces, the phase shifts vanish if the
momentum  of a scattered particle tends to zero \cite{LandauIII}.
In our case, the nonzero value of $\delta_{m n}\left(0\right)$ is caused by the
long-range ($\propto \rho^{-1}$) character of gauge field (\ref{II:7}).

In the following, we discuss this issue in more detail.
Since the  phase  shifts depend on the dimensionless combination $\tilde{k} = k
\lambda$, the regime  of  small  $k$  is  equivalent  to  the  regime  of small
$\lambda$.
Due to the long-range character  of  the  gauge  field  $A_{\mu}$, the function
$A_{m n}(\rho)$ included in the system of differential equations (\ref{III:6a})
and (\ref{III:6b}) tends to a  constant value $m-n$ as $\lambda \rightarrow 0$.
As  a  result,  the  system   of   differential   equations  (\ref{III:6a}) and
(\ref{III:6b}) is simplified,  and  its  solution  can be expressed in terms of
Bessel functions $J_{m - n \pm 1/2}(k \rho)$.
The free motion of fermions corresponds to $n =0$ (the absence of a $\mathbb{CP
}^{N-1}$ soliton).
In this case, the  function $A_{m 0}(\rho) = m$, and the solution to the system
of differential  equations  (\ref{III:6a})  and  (\ref{III:6b}) is expressed in
terms of Bessel functions $J_{m \pm 1/2}(k \rho)$.
Using the well-known asymptotic  expansions of the Bessel functions, it is easy
to show that the phase  shift between $J_{m - n \pm 1/2}(k \rho)$ and $J_{m \pm
1/2}(k \rho)$ is $\pi n /2$.
This can be regarded as the phase shift at zero $\tilde{k}$, and can be written
as $\text{sign}\left(n\right) \pi/2$ for $n = \pm 1$.
This expression is compatible  with  the result $\delta_{m n}\left( 0 \right) =
\text{sign}\left(mn\right)\pi/2$, as the phase shifts are defined modulo $\pi$.

Using analytical and numerical  methods, we were able to establish the behavior
of the phase shifts $\delta_{m n}$ in the region of small $\tilde{k}$ as
\begin{equation}
\delta_{m n}(\tilde{k}) \approx \left\{
\begin{array}{c}
\dfrac{\pi }{2}+\dfrac{\pi}{\ln (\tilde{k}^{2})}
\qquad\qquad\, \text{if}\quad m n = \dfrac{1}{2} \\
s_{m n} \dfrac{\pi }{2}+\alpha_{m n}\tilde{k}^{2\beta_{m n}}
\quad \text{if} \quad m n \neq \dfrac{1}{2}
\end{array}
\right.,                                                            \label{V:2}
\end{equation}
where $\tilde{k} = k \lambda$,  $s_{m n}  =  \text{sign}(m n)$,  $\beta_{m n} =
\left\vert m\right\vert + 1/2 - s_{m n}$,  and  $\alpha_{m n}$ are coefficients
satisfying the condition $\alpha_{m n} = \alpha_{-m\,-n}$.
Based on Eqs.~(\ref{III:28}), (\ref{III:32}), and (\ref{V:2}), we can write the
corresponding expressions for the partial amplitudes as
\begin{equation}
f_{m n} \approx \left\{
\begin{array}{c}
\sqrt{\dfrac{2}{\pi k}}\left[i-\dfrac{2}{\ln(\tilde{k}^{2})}\right]
\qquad \text{if} \quad m n = 1/2 \\
\sqrt{\dfrac{2}{\pi k}}\left[i-\alpha_{m n}\tilde{k}^{2\beta_{m n}}\right]
\;\;\text{if} \quad m n \neq 1/2
\end{array}
\right.                                                             \label{V:3}
\end{equation}
We see that as $\tilde{k} \rightarrow  0$,  all  partial amplitudes tend to the
same limiting form $i \sqrt{2/(\pi k)}$.
Accordingly, the partial cross-sections $\sigma_{m n}= 2 \pi \left\vert f_{m n}
\right\vert^{2}$ tend to $4 k^{-1}$, and hence attain the unitary bound in this
limit.
Note that  all  partial  waves  make  the  same  contribution  to  the  fermion
scattering when $\tilde{k} = k \lambda \rightarrow 0$.
This is due to the fact that $\left\vert\delta_{m n}(0)\right\vert = \pi/2$ for
all $m$.

Let us define $\tilde{k}_{1/2}$ as  the  value  of  $\tilde{k}$ at which $\left
\vert \delta_{m n}\right \vert$ takes  the  value of $\pi/4$, i.e., half of the
maximum value $\pi/2$.
We have established numerically that  the dependence of the parameter $\tilde{k
}_{1/2}$ on the eigenvalue  $m$  of  the  angular  momentum has the approximate
linear form
\begin{equation}
\tilde{k}_{1/2} = k_{1/2} \lambda
\approx 1.74 \left\vert m \right\vert.                              \label{V:4}
\end{equation}
As $\left\vert m \right\vert$  grows,  the  main  contribution  to  the angular
momentum in Eq.~(\ref{III:4}) comes from its orbital part.
Eq.~(\ref{V:4}) then tells us that the  orbital part of the angular momentum is
approximately proportional to  the  fermion  momentum  $k_{1/2}$ and the linear
size $\lambda$ of  the soliton, which is consistent with classical conceptions.

It follows from Figs.~\ref{fig1}  and  \ref{fig2} that  for  fixed $\tilde{k}$,
the absolute values of $\delta_{m n}$ increase  with an increase in $\left\vert
m \right\vert$.
This is true for both positive and negative $m$.
Using the formula $\sigma_{m n} = 4 k^{-1}\sin^{2}(\delta_{m  n})$, we conclude
that the partial cross-sections $\sigma_{m n}$ behave similarly.
We see that for all values of $k$,  the  contribution  of  partial waves to the
fermion-soliton scattering increases with an  increase  in  $\left\vert m\right
\vert$.
This is because long-range  gauge field (\ref{II:7}) of the $\mathbb{CP}^{N-1}$
soliton makes a  significant  contribution  to  the fermion scattering, even at
large distances from the soliton.
As  $\tilde{k}$  increases,  $\left\vert  \delta_{m  n}  \right\vert$ decreases
monotonically and tends to zero as $\tilde{k} \rightarrow \infty$.
We have found numerically that in this limit, the phase shifts
\begin{equation}
\delta_{m n}\bigl(\tilde{k}\bigr) \approx \frac{\pi}{2\tilde{k}}
\left(n m  - \frac{1}{4}\right).                                    \label{V:5}
\end{equation}

These features of the curves $\delta_{m n}(\tilde{k})$ can be understood in the
framework of the quasiclassical approximation.
Using methods of  scattering  theory \cite{LandauIII}, it can be shown that for
sufficiently   large   $\left\vert   m   \right\vert$   and   $\tilde{k}$,  the
fermion-soliton scattering is quasiclassical.
There is an approximate quasiclassical  expression  for the phase shifts, which
in our case can be written as
\begin{eqnarray}
\delta_{mn}\bigl(\tilde{k}\bigr)  &\approx &\int\limits_{\tilde{
\rho }_{0}}^{\infty }\left[\tilde{k}^{2}-\frac{\left(m-1/2\right)^{2}
}{\tilde{\rho }^{2}}-W\left(\tilde{\rho},m,n\right) \right]^{1/2}
d\tilde{\rho}                                                    \nonumber
 \\
&&-\int\limits_{\tilde{\rho}_{0}}^{\infty }\left[\tilde{k}^{2}-
\frac{\left( m-1/2\right) ^{2}}{\tilde{\rho }^{2}}\right]^{1/2}
d\tilde{\rho},                                                      \label{V:6}
\end{eqnarray}
where the potential
\begin{equation}
W\left(\tilde{\rho },m,n\right) = \frac{n\left(n - 2m + 1\right)}{1+
\tilde{\rho}^{2}}-\frac{n\left(2 + n\right)}{\left(1+\tilde{\rho }
^{2}\right) ^{2}}                                                   \label{V:7}
\end{equation}
and the lower limit of integration
\begin{equation}
\tilde{\rho}_{0}\approx \left\vert m-1/2\right\vert\tilde{k}^{-1}.  \label{V:8}
\end{equation}
If quasiclassical conditions are  fulfilled,  then  the  potential $W$  will be
small compared to the term $\tilde{k}^{2} - \left(m - 1/2\right)^{2}\tilde{\rho
}^{-2}$ in the region making the  main  contribution  to  the first integral in
Eq.~(\ref{V:6}).
Expanding the integrand of  the  first  integral  in $W$  and keeping the first
expansion term, we can  obtain  an  approximate  analytical  expression for the
phase shifts as
\begin{eqnarray}
\delta_{mn}\bigl( \tilde{k}\bigr)  &\approx &\frac{\pi n}{2}\left[
(2m-n-1)\left( 2\tilde{k}^{2}+(2m-1)^{2}\right) \right.             \nonumber
  \\
&&\left. +2\tilde{k}^{2}(2m+1)\right]\!\!\left[4\tilde{k}
^{2}+(2m-1)^{2}\right] ^{-\frac{3}{2}}\!\!.                         \label{V:9}
\end{eqnarray}
From Eq.~(\ref{V:9}), we can obtain  two  asymptotic  expressions for the phase
shifts.
The first  is  valid  for  $\tilde{k}  \rightarrow  \infty$  and fixed $m$, and
coincides with Eq.~(\ref{V:5}).
The second is valid for $\left\vert m \right\vert \rightarrow \infty$ and fixed
$\tilde{k}$, and has the form
\begin{equation}
\delta_{mn}\bigl( \tilde{k}\bigr) \sim s_{m n}
\frac{\pi}{2} - \frac{\pi}{4\left\vert m\right\vert},              \label{V:10}
\end{equation}
where the factor $s_{m n} = \text{sign}(m n)$.
Thus, the asymptotic behavior in  Eq.~(\ref{V:5}) can  be  obtained  within the
quasiclassical approximation.
Furthermore, Eq.~(\ref{V:10}) tells  us  that  for fixed $\tilde{k}$, the phase
shifts $\delta_{mn}\bigl(\tilde{k}\bigr)\rightarrow s_{mn}\pi/2$ as $\left\vert
m \right\vert \rightarrow  \infty$,  which  is  consistent  with  the numerical
results.
It follows  that  the  partial  cross-sections $\sigma_{m n} = 4 k^{-1}\sin^{2}
(\delta_{m n})$ reach the unitary bound $4 k^{-1}$ as $\left\vert m \right\vert
\rightarrow \infty$.

As already noted, this behavior of the phase shifts $\delta_{mn}(\tilde{k})$ is
due to the slow ($\propto \rho^{-1}$) decrease  of gauge field (\ref{II:7}) far
from the soliton.
This behavior of the gauge field  leads  to  the long-range asymptotics $W \sim
n (n - 2m - 1)\tilde{\rho}^{-2}$   of  quasiclassical   potential  (\ref{V:7}).
It is this asymptotic behavior of $W$  that  leads  to  the fact that the phase
shifts $\delta_{m n}\bigl(\tilde{k}\bigr)\rightarrow s_{m n}\pi /2$  as  $\left
\vert m \right\vert \rightarrow \infty$.
Indeed, a faster decrease in gauge field  (\ref{II:7}) 
leads to a faster decrease in quasiclassical potential (\ref{V:7}).
It can be shown, however, that if  the  quasiclassical  potential $W$ decreases
more rapidly than $\tilde{\rho}^{-2}$,  quasiclassical  phase shift (\ref{V:6})
tends to zero as $\left\vert m \right\vert \rightarrow \infty$.
In this case, the contribution of partial  waves with sufficiently large $\left
\vert m \right\vert$ to the fermion scattering becomes negligibly small.

It follows from the results obtained that the difference in the phase shifts is
\begin{equation}
\delta_{m n}\left(0\right) - \delta_{mn}\left( \infty \right)
= s_{mn}\pi /2.                                                    \label{V:11}
\end{equation}
This contradicts Levinson's theorem \cite{levinson_49}, according to which this
difference must be equal to $\pi$ multiplied  by  the number of bound fermionic
states in the partial channel with given values of $m$ and $n$.
Since there are no bound fermionic states in our case, the difference $\delta_{
m n}\left(0\right)-\delta_{mn}\left(\infty\right)$ must be equal to zero, which
contradicts Eq.~(\ref{V:11}).
The reason for this is that  one  of  the  conditions  for the applicability of
Levinson's theorem is a  rather fast  decrease (faster than $\rho^{-3}$) in the
potential term at infinity \cite{Taylor}.
In our  case,  the  slow  decrease  ($\sim \! \rho^{-2}$)  of  the potential in
Eq.~(\ref{V:7}) for large $\rho$ makes Levinson's theorem inapplicable.

It follows from the results in Sec.~\ref{sec:IV} that in the Born approximation,
the partial amplitudes $f_{m n}$ are real.
At the same time, Eqs.~(\ref{III:28}) and (\ref{III:32})  tell  us that $\text{
Im}\left[ f_{m n}\right] =\sqrt{\pi k/2}\left\vert f_{m n}\right\vert^{2} > 0$.
We see that according to  scattering theory \cite{LandauIII, Taylor}, unitarity
is broken in the Born approximation.
However, it follows  from  Eqs.~(\ref{III:28}), (\ref{III:32}), and (\ref{V:5})
that $\text{Im} \left[f_{m n} \right]  \sim  \pi^{2} \left(1 - 4 m n\right)^{2}
(32\tilde{k}^{2})^{-1}\left(2 \pi k\right)^{-1/2}$,  and   hence  tends to zero
$\propto k^{-5/2}$ as $k \rightarrow \infty$.
Consequently, the Born approximation  becomes applicable in the region of large
fermion momenta $k$.

\begin{figure}[tbp]
\includegraphics[width=0.5\textwidth]{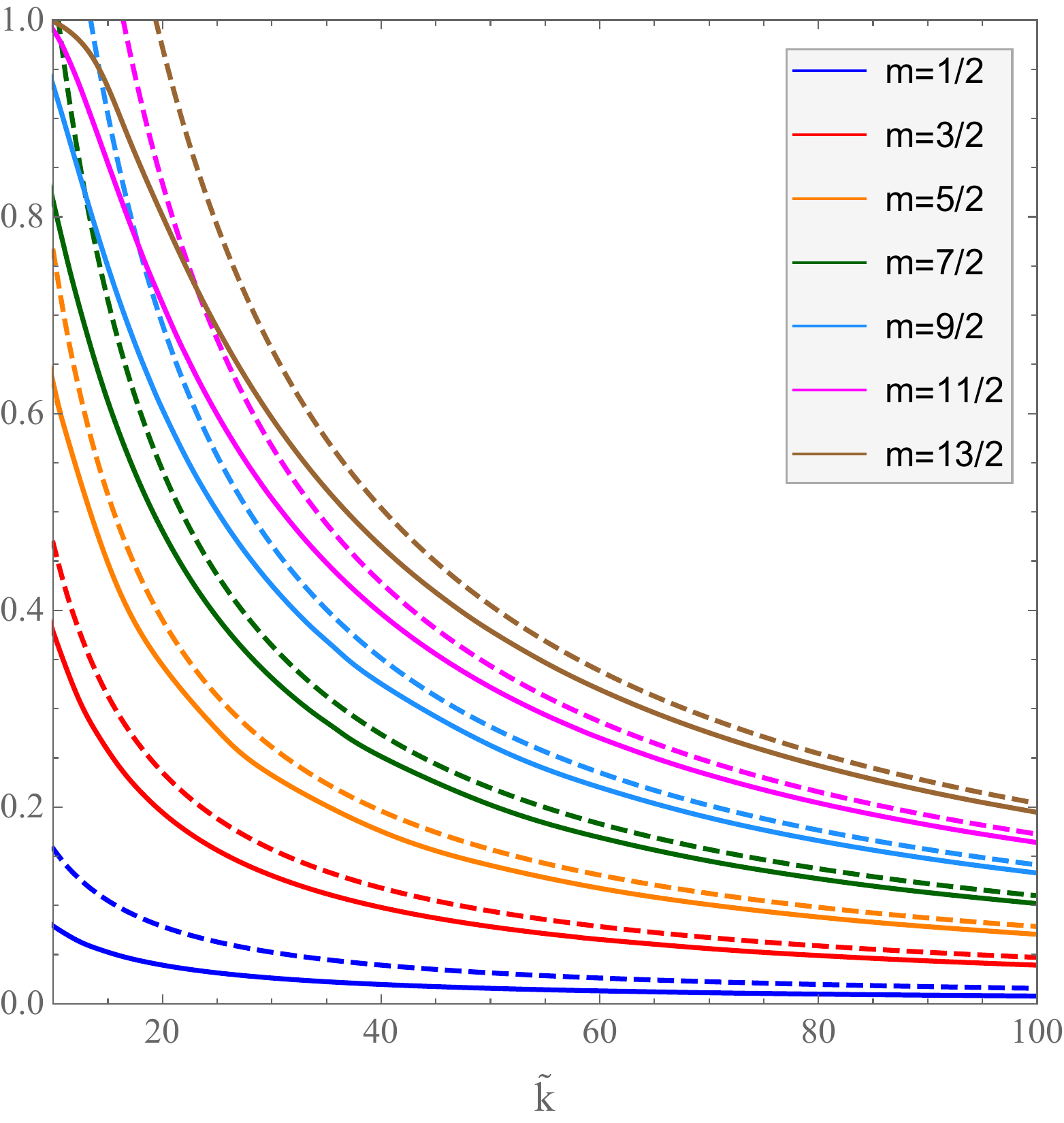}
\caption{\label{fig3}  Dependence of $\sqrt{2\pi k}\,\text{Re}[f_{m n}]$ on the
dimensionless combination $\tilde{k} =  k\lambda$ for $m = 1/2,\, 3/2,\, 5/2,\,
7/2,\, 9/2,\, 11/2,\, 13/2$, and $n = 1$.  The solid curves correspond to exact
solution  (\ref{III:18}),  and   the  dashed   curves  correspond  to  the Born
approximation (\ref{IV:3}).}
\end{figure}

\begin{figure}[tbp]
\includegraphics[width=0.5\textwidth]{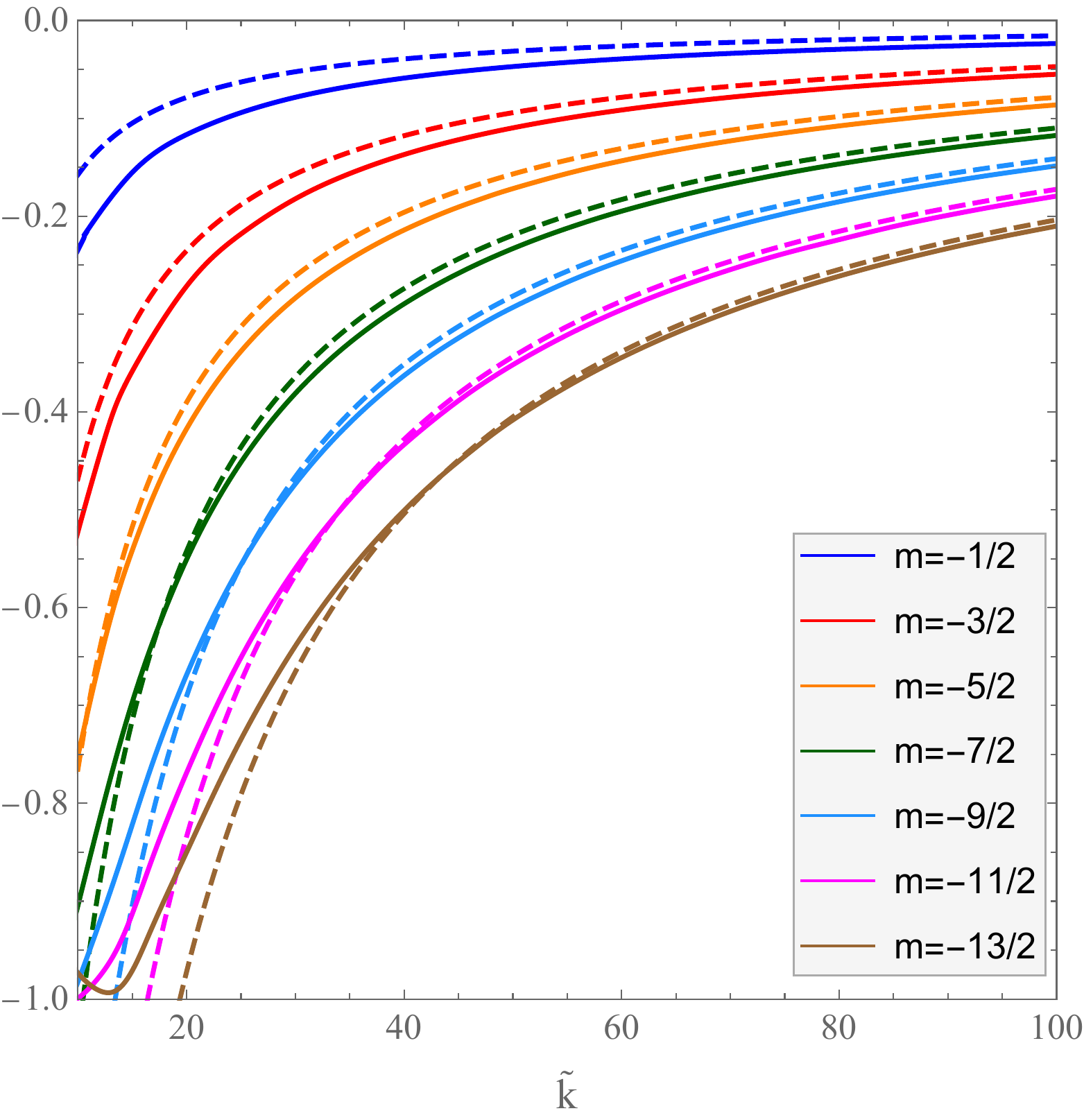}
\caption{\label{fig4}  Dependence of $\sqrt{2\pi k}\,\text{Re}[f_{m n}]$ on the
dimensionless combination $\tilde{k} =  k\lambda$ for $m =-1/2,\,-3/2,\,-5/2,\,
-7/2,\,-9/2,\,-11/2,\,-13/2$, and $n = 1$. The solid curves correspond to exact
solution  (\ref{III:18}),  and   the  dashed   curves  correspond  to  the Born
approximation (\ref{IV:3}).}
\end{figure}

It was shown in Sec.~\ref{sec:III} that the  phase shifts $\delta_{m n}$ depend
only on the dimensionless combination $\tilde{k} = k \lambda$.
Eqs.~(\ref{III:28}) and (\ref{III:32})  then  tell  us  that  the dimensionless
combinations $\sqrt{2 \pi k} f_{m n}$ also depend only on $\tilde{k}$.
Figure~\ref{fig3} shows the dependences of $\sqrt{2\pi k}\text{Re}[f_{m n}]$
on  $\tilde{k}$ for  the  first  few  positive  eigenvalues  $m$ of the angular
momentum and  the  soliton  winding number $n = 1$.
In Fig.~\ref{fig3}, the solid curves correspond to exact solution (\ref{III:18})
and the dashed curves correspond to the Born approximation (\ref{IV:3}).
Similar curves for  negative  eigenvalues  $m$  are  shown  in Fig.~\ref{fig4}.

From Figs.~\ref{fig3} and \ref{fig4}, it follows  that the accuracy of the Born
approximation improves with an increase in $\tilde{k}$.
At the same time, a comparison of Eqs.~(\ref{IV:11}) and (\ref{V:5}) shows that
even for  large  $\tilde{k}$,  the  Born  phases  differ  from  those  obtained
numerically  (or  within  the  quasiclassical  approximation)  by  a  shift  of
$\pi/(8\tilde{k})$.
This difference is due to the violation of unitarity in the Born approximation,
and becomes  insignificant  with   an  increase  in $\left\vert m \right\vert$.
Note that  the  Born  partial  amplitudes  change  sign  under  the replacement
$m \rightarrow -m$, which is a consequence of Eq.~(\ref{IV:8}).
This property, however, is true  only  in  the  Born approximation, and is lost
when we pass to the exact partial amplitudes.
Instead, the exact partial amplitudes satisfy the condition $f_{m n}(k) = f_{-m
\, -n}(k)$, which is a consequence of general symmetry relation (\ref{III:33}).

\section{\label{sec:VI} Conclusion}

In this paper,  we have investigated fermion scattering on topological solitons
of the $(2+1)$-dimensional $\mathbb{CP}^{N-1}$  model  in  the framework of the
background field approximation.
In particular, we   found  exact  solutions  to  the  Dirac equation describing
fermionic states in the background fields of  $\mathbb{CP}^{N-1}$ solitons with
winding numbers $n = \pm 1$.
It turns out that  these  exact  solutions  can  be  expressed  in terms of the
confluent Heun functions.
The  symmetry  properties  of  the  fermionic  wave  functions  under  discrete
transformations of the Dirac equation were found, which allowed us to establish
the discrete symmetry property of the phase shifts.
We studied the the presence  of fermionic bound states in the background fields
of the $\mathbb{CP}^{N-1}$ solitons with winding  numbers $n = \pm 1$, and came
to the conclusion that there are no such states.

Within the framework  of  the  background  field  approximation, the process of
fermion-soliton scattering is elastic,  and can therefore be fully described in
terms of phase shifts.
However, the absence of analytical asymptotics for the confluent Heun functions
makes it impossible to obtain analytical expressions for the phase shifts.
In view  of  this,  we  studied  the  fermion-soliton  scattering  in  the Born
approximation,  which  gave   us  the  opportunity  to  obtain  an  approximate
analytical  expressions  for  the  phase   shifts,  scattering  amplitudes, and
differential cross-sections, and to study their asymptotic forms.
We found  that  the  total  cross-section  of  the  fermion-soliton  scattering
diverges due to the long-range character of the soliton field.
However, the transport cross-section  of  the  fermion-soliton scattering turns
out to be finite, and can be expressed  in  terms  of the Meijer $G$-functions.

We  have  also  performed  a  numerical  study  of  fermion  scattering  in the
background fields of  the $\mathbb{CP}^{N - 1}$  solitons  with winding numbers
$n = \pm 1$.
In particular, it was found that the phase shifts $\delta_{m n}$ depend only on
the dimensionless combination $\tilde{k}=k\lambda$, and the curves $\delta_{m\,
\pm 1}(\tilde{k})$  were  obtained  for  $\left \vert m \right \vert \le 13/2$.
The main feature of the  curves  $\delta_{m n}(\tilde{k})$ is that they tend to
a non-zero  value $\delta_{m n}\left( 0 \right) = \text{sign} \left(m n \right)
\pi/2$ as $\tilde{k} \rightarrow 0$.
At the same  time,  the  phase  shifts  $\delta_{m n}(\tilde{k})$  tend to zero
$\propto \tilde{k}^{-1}$ as $\tilde{k} \rightarrow \infty$.
The nonzero value of the difference $\delta_{m n}(0) - \delta_{m n}(\infty)$ in
spite of the absence of  bound  fermionic states is related to long-range gauge
field (\ref{II:7}) of the $\mathbb{CP}^{N-1}$ soliton.

We have found that as $\left\vert m\right\vert$ increases, the curves $\delta_{
m n}(\tilde{k})$ shift to the region of larger $\tilde{k}$.
Using the quasiclassical approximation,  we  have  shown  that the phase shifts
$\delta_{m n}\bigl(\tilde{k}\bigr)$  tend  to $\text{sign}(m n)\pi/2$ as $\left
\vert m \right\vert $  tends  to  $ \infty$,   which  is  consistent  with  our
numerical results.
It follows that partial waves with arbitrarily large $\left\vert m \right\vert$
make  a  significant  contribution  to   fermion  scattering  at  any  value of
$\tilde{k}$ (including small values).
This feature of fermion-soliton scattering  is  also  related to the long-range
character of the gauge field in Eq.~(\ref{II:7}).




\bibliography{article}

\providecommand{\noopsort}[1]{}\providecommand{\singleletter}[1]{#1}%
\begin{thebibliography}{37}%
\makeatletter
\providecommand \@ifxundefined [1]{%
 \@ifx{#1\undefined}
}%
\providecommand \@ifnum [1]{%
 \ifnum #1\expandafter \@firstoftwo
 \else \expandafter \@secondoftwo
 \fi
}%
\providecommand \@ifx [1]{%
 \ifx #1\expandafter \@firstoftwo
 \else \expandafter \@secondoftwo
 \fi
}%
\providecommand \natexlab [1]{#1}%
\providecommand \enquote  [1]{``#1''}%
\providecommand \bibnamefont  [1]{#1}%
\providecommand \bibfnamefont [1]{#1}%
\providecommand \citenamefont [1]{#1}%
\providecommand \href@noop [0]{\@secondoftwo}%
\providecommand \href [0]{\begingroup \@sanitize@url \@href}%
\providecommand \@href[1]{\@@startlink{#1}\@@href}%
\providecommand \@@href[1]{\endgroup#1\@@endlink}%
\providecommand \@sanitize@url [0]{\catcode `\\12\catcode `\$12\catcode
  `\&12\catcode `\#12\catcode `\^12\catcode `\_12\catcode `\%12\relax}%
\providecommand \@@startlink[1]{}%
\providecommand \@@endlink[0]{}%
\providecommand \url  [0]{\begingroup\@sanitize@url \@url }%
\providecommand \@url [1]{\endgroup\@href {#1}{\urlprefix }}%
\providecommand \urlprefix  [0]{URL }%
\providecommand \Eprint [0]{\href }%
\providecommand \doibase [0]{https://doi.org/}%
\providecommand \selectlanguage [0]{\@gobble}%
\providecommand \bibinfo  [0]{\@secondoftwo}%
\providecommand \bibfield  [0]{\@secondoftwo}%
\providecommand \translation [1]{[#1]}%
\providecommand \BibitemOpen [0]{}%
\providecommand \bibitemStop [0]{}%
\providecommand \bibitemNoStop [0]{.\EOS\space}%
\providecommand \EOS [0]{\spacefactor3000\relax}%
\providecommand \BibitemShut  [1]{\csname bibitem#1\endcsname}%
\let\auto@bib@innerbib\@empty
\bibitem [{\citenamefont {Manton}\ and\ \citenamefont
  {Sutclffe}(2004)}]{Manton}%
  \BibitemOpen
  \bibfield  {author} {\bibinfo {author} {\bibfnamefont {N.}~\bibnamefont
  {Manton}}\ and\ \bibinfo {author} {\bibfnamefont {P.}~\bibnamefont
  {Sutclffe}},\ }\href@noop {} {\emph {\bibinfo {title} {Topological
  Solitons}}}\ (\bibinfo  {publisher} {Cambridge University Press},\ \bibinfo
  {address} {Cambridge},\ \bibinfo {year} {2004})\BibitemShut {NoStop}%
\bibitem [{\citenamefont {Weinberg}(2012)}]{E_Weinberg}%
  \BibitemOpen
  \bibfield  {author} {\bibinfo {author} {\bibfnamefont {E.~J.}\ \bibnamefont
  {Weinberg}},\ }\href@noop {} {\emph {\bibinfo {title} {Classical Solutions in
  Quantum Field Theory: Solitons and Instantons in High Energy Physics}}}\
  (\bibinfo  {publisher} {Cambridge University Press},\ \bibinfo {address}
  {Cambridge},\ \bibinfo {year} {2012})\BibitemShut {NoStop}%
\bibitem [{\citenamefont {Zakrzewski}(1989)}]{Zakrzewski}%
  \BibitemOpen
  \bibfield  {author} {\bibinfo {author} {\bibfnamefont {W.~J.}\ \bibnamefont
  {Zakrzewski}},\ }\href@noop {} {\emph {\bibinfo {title} {Low Dimensional
  Sigma Models}}}\ (\bibinfo  {publisher} {Taylor \& Francis},\ \bibinfo
  {address} {London},\ \bibinfo {year} {1989})\BibitemShut {NoStop}%
\bibitem [{\citenamefont {Abrikosov}(1957)}]{abr}%
  \BibitemOpen
  \bibfield  {author} {\bibinfo {author} {\bibfnamefont {A.~A.}\ \bibnamefont
  {Abrikosov}},\ }\href@noop {} {\bibfield  {journal} {\bibinfo  {journal} {Zh.
  Exp. Teor. Fiz.}\ }\textbf {\bibinfo {volume} {32}},\ \bibinfo {pages} {1442}
  (\bibinfo {year} {1957})},\ \translation{Sov. Phys. JETP \textbf{5}, 1174
  (1957)}\BibitemShut {NoStop}%
\bibitem [{\citenamefont {Nielsen}\ and\ \citenamefont
  {Olesen}(1973)}]{nielsen}%
  \BibitemOpen
  \bibfield  {author} {\bibinfo {author} {\bibfnamefont {H.~B.}\ \bibnamefont
  {Nielsen}}\ and\ \bibinfo {author} {\bibfnamefont {P.}~\bibnamefont
  {Olesen}},\ }\href@noop {} {\bibfield  {journal} {\bibinfo  {journal} {Nucl.
  Phys. B}\ }\textbf {\bibinfo {volume} {61}},\ \bibinfo {pages} {45} (\bibinfo
  {year} {1973})}\BibitemShut {NoStop}%
\bibitem [{\citenamefont {Belavin}\ and\ \citenamefont {Polyakov}(1975)}]{BP}%
  \BibitemOpen
  \bibfield  {author} {\bibinfo {author} {\bibfnamefont {A.~A.}\ \bibnamefont
  {Belavin}}\ and\ \bibinfo {author} {\bibfnamefont {A.~M.}\ \bibnamefont
  {Polyakov}},\ }\href@noop {} {\bibfield  {journal} {\bibinfo  {journal}
  {Pis'ma Zh. Exp. Teor. Fiz.}\ }\textbf {\bibinfo {volume} {22}},\ \bibinfo
  {pages} {503} (\bibinfo {year} {1975})},\ \translation{JETP Lett.
  \textbf{22}, 245 (1975)}\BibitemShut {NoStop}%
\bibitem [{\citenamefont {Cremmer}\ and\ \citenamefont
  {Scherk}(1978)}]{cremmer}%
  \BibitemOpen
  \bibfield  {author} {\bibinfo {author} {\bibfnamefont {E.}~\bibnamefont
  {Cremmer}}\ and\ \bibinfo {author} {\bibfnamefont {J.}~\bibnamefont
  {Scherk}},\ }\href@noop {} {\bibfield  {journal} {\bibinfo  {journal} {Phys.
  Lett. B}\ }\textbf {\bibinfo {volume} {74}},\ \bibinfo {pages} {341}
  (\bibinfo {year} {1978})}\BibitemShut {NoStop}%
\bibitem [{\citenamefont {Eichenherr}(1978)}]{eichenherr}%
  \BibitemOpen
  \bibfield  {author} {\bibinfo {author} {\bibfnamefont {H.}~\bibnamefont
  {Eichenherr}},\ }\href@noop {} {\bibfield  {journal} {\bibinfo  {journal}
  {Nucl. Phys. B}\ }\textbf {\bibinfo {volume} {146}},\ \bibinfo {pages} {215}
  (\bibinfo {year} {1978})}\BibitemShut {NoStop}%
\bibitem [{\citenamefont {Golo}\ and\ \citenamefont
  {Perelomov}(1978{\natexlab{a}})}]{per1}%
  \BibitemOpen
  \bibfield  {author} {\bibinfo {author} {\bibfnamefont {V.~L.}\ \bibnamefont
  {Golo}}\ and\ \bibinfo {author} {\bibfnamefont {A.~M.}\ \bibnamefont
  {Perelomov}},\ }\href@noop {} {\bibfield  {journal} {\bibinfo  {journal}
  {Lett. Math. Phys.}\ }\textbf {\bibinfo {volume} {2}},\ \bibinfo {pages}
  {477} (\bibinfo {year} {1978}{\natexlab{a}})}\BibitemShut {NoStop}%
\bibitem [{\citenamefont {Golo}\ and\ \citenamefont
  {Perelomov}(1978{\natexlab{b}})}]{per2}%
  \BibitemOpen
  \bibfield  {author} {\bibinfo {author} {\bibfnamefont {V.~L.}\ \bibnamefont
  {Golo}}\ and\ \bibinfo {author} {\bibfnamefont {A.~M.}\ \bibnamefont
  {Perelomov}},\ }\href@noop {} {\bibfield  {journal} {\bibinfo  {journal}
  {Phys. Lett. B}\ }\textbf {\bibinfo {volume} {79}},\ \bibinfo {pages} {112}
  (\bibinfo {year} {1978}{\natexlab{b}})}\BibitemShut {NoStop}%
\bibitem [{\citenamefont {D'Adda}\ \emph {et~al.}(1978)\citenamefont {D'Adda},
  \citenamefont {Luscher},\ and\ \citenamefont {Vecchia}}]{dadda}%
  \BibitemOpen
  \bibfield  {author} {\bibinfo {author} {\bibfnamefont {A.}~\bibnamefont
  {D'Adda}}, \bibinfo {author} {\bibfnamefont {M.}~\bibnamefont {Luscher}},\
  and\ \bibinfo {author} {\bibfnamefont {P.~D.}\ \bibnamefont {Vecchia}},\
  }\href@noop {} {\bibfield  {journal} {\bibinfo  {journal} {Nucl. Phys. B}\
  }\textbf {\bibinfo {volume} {146}},\ \bibinfo {pages} {63} (\bibinfo {year}
  {1978})}\BibitemShut {NoStop}%
\bibitem [{\citenamefont {Witten}(1979)}]{witten}%
  \BibitemOpen
  \bibfield  {author} {\bibinfo {author} {\bibfnamefont {E.}~\bibnamefont
  {Witten}},\ }\href@noop {} {\bibfield  {journal} {\bibinfo  {journal} {Nucl.
  Phys. B}\ }\textbf {\bibinfo {volume} {149}},\ \bibinfo {pages} {285}
  (\bibinfo {year} {1979})}\BibitemShut {NoStop}%
\bibitem [{\citenamefont {Polyakov}(1975)}]{polyakov2}%
  \BibitemOpen
  \bibfield  {author} {\bibinfo {author} {\bibfnamefont {A.~M.}\ \bibnamefont
  {Polyakov}},\ }\href@noop {} {\bibfield  {journal} {\bibinfo  {journal}
  {Phys. Lett. B}\ }\textbf {\bibinfo {volume} {59}},\ \bibinfo {pages} {79}
  (\bibinfo {year} {1975})}\BibitemShut {NoStop}%
\bibitem [{\citenamefont {Hanany}\ and\ \citenamefont
  {Tong}({\natexlab{a}})}]{hanany_2003}%
  \BibitemOpen
  \bibfield  {author} {\bibinfo {author} {\bibfnamefont {A.}~\bibnamefont
  {Hanany}}\ and\ \bibinfo {author} {\bibfnamefont {D.}~\bibnamefont {Tong}},\
  }\href@noop {} {\bibfield  {journal} {\bibinfo  {journal} {J. High Energy
  Phys.}\ }\textbf {\bibinfo {volume} {\textmd{07 (2003) 037}}}}\BibitemShut
  {NoStop}%
\bibitem [{\citenamefont {Auzzi}\ \emph {et~al.}(2003)\citenamefont {Auzzi},
  \citenamefont {Bolognesi}, \citenamefont {Evslin}, \citenamefont {Konishi},\
  and\ \citenamefont {Yung}}]{auzzi_2003}%
  \BibitemOpen
  \bibfield  {author} {\bibinfo {author} {\bibfnamefont {R.}~\bibnamefont
  {Auzzi}}, \bibinfo {author} {\bibfnamefont {S.}~\bibnamefont {Bolognesi}},
  \bibinfo {author} {\bibfnamefont {J.}~\bibnamefont {Evslin}}, \bibinfo
  {author} {\bibfnamefont {K.}~\bibnamefont {Konishi}},\ and\ \bibinfo {author}
  {\bibfnamefont {A.}~\bibnamefont {Yung}},\ }\href@noop {} {\bibfield
  {journal} {\bibinfo  {journal} {Nucl. Phys. B}\ }\textbf {\bibinfo {volume}
  {673}},\ \bibinfo {pages} {187} (\bibinfo {year} {2003})}\BibitemShut
  {NoStop}%
\bibitem [{\citenamefont {Shifman}\ and\ \citenamefont
  {Yung}(2004)}]{shifman_2003}%
  \BibitemOpen
  \bibfield  {author} {\bibinfo {author} {\bibfnamefont {M.}~\bibnamefont
  {Shifman}}\ and\ \bibinfo {author} {\bibfnamefont {A.}~\bibnamefont {Yung}},\
  }\href@noop {} {\bibfield  {journal} {\bibinfo  {journal} {Phys. Rev. D}\
  }\textbf {\bibinfo {volume} {70}},\ \bibinfo {pages} {045004} (\bibinfo
  {year} {2004})}\BibitemShut {NoStop}%
\bibitem [{\citenamefont {Hanany}\ and\ \citenamefont
  {Tong}({\natexlab{b}})}]{hanany_2004}%
  \BibitemOpen
  \bibfield  {author} {\bibinfo {author} {\bibfnamefont {A.}~\bibnamefont
  {Hanany}}\ and\ \bibinfo {author} {\bibfnamefont {D.}~\bibnamefont {Tong}},\
  }\href@noop {} {\bibfield  {journal} {\bibinfo  {journal} {J. High Energy
  Phys.}\ }\textbf {\bibinfo {volume} {\textmd{04 (2004) 066}}}}\BibitemShut
  {NoStop}%
\bibitem [{\citenamefont {Shifman}\ and\ \citenamefont
  {Yung}(2007)}]{shif_yung}%
  \BibitemOpen
  \bibfield  {author} {\bibinfo {author} {\bibfnamefont {M.}~\bibnamefont
  {Shifman}}\ and\ \bibinfo {author} {\bibfnamefont {A.}~\bibnamefont {Yung}},\
  }\href@noop {} {\bibfield  {journal} {\bibinfo  {journal} {Rev. Mod. Phys.}\
  }\textbf {\bibinfo {volume} {79}},\ \bibinfo {pages} {1139} (\bibinfo {year}
  {2007})}\BibitemShut {NoStop}%
\bibitem [{\citenamefont {Tong}(2009)}]{tong}%
  \BibitemOpen
  \bibfield  {author} {\bibinfo {author} {\bibfnamefont {D.}~\bibnamefont
  {Tong}},\ }\href@noop {} {\bibfield  {journal} {\bibinfo  {journal} {Ann.
  Phys. (N. Y.)}\ }\textbf {\bibinfo {volume} {324}},\ \bibinfo {pages} {30}
  (\bibinfo {year} {2009})}\BibitemShut {NoStop}%
\bibitem [{\citenamefont {Tsvelik}(1995)}]{Tsvelik}%
  \BibitemOpen
  \bibfield  {author} {\bibinfo {author} {\bibfnamefont {A.~M.}\ \bibnamefont
  {Tsvelik}},\ }\href@noop {} {\emph {\bibinfo {title} {Quantum Field Theory in
  Condensed Matter}}}\ (\bibinfo  {publisher} {Cambridge University Press},\
  \bibinfo {address} {Cambridge},\ \bibinfo {year} {1995})\BibitemShut
  {NoStop}%
\bibitem [{\citenamefont {Mottola}\ and\ \citenamefont
  {Wipf}(1989)}]{mottola_1989}%
  \BibitemOpen
  \bibfield  {author} {\bibinfo {author} {\bibfnamefont {E.}~\bibnamefont
  {Mottola}}\ and\ \bibinfo {author} {\bibfnamefont {A.}~\bibnamefont {Wipf}},\
  }\href@noop {} {\bibfield  {journal} {\bibinfo  {journal} {Phys. Rev. D}\
  }\textbf {\bibinfo {volume} {39}},\ \bibinfo {pages} {588} (\bibinfo {year}
  {1989})}\BibitemShut {NoStop}%
\bibitem [{\citenamefont {Abdalla}\ \emph {et~al.}(1982)\citenamefont
  {Abdalla}, \citenamefont {Abdalla},\ and\ \citenamefont
  {Gomes}}]{abdalla_1982}%
  \BibitemOpen
  \bibfield  {author} {\bibinfo {author} {\bibfnamefont {E.}~\bibnamefont
  {Abdalla}}, \bibinfo {author} {\bibfnamefont {M.}~\bibnamefont {Abdalla}},\
  and\ \bibinfo {author} {\bibfnamefont {M.}~\bibnamefont {Gomes}},\
  }\href@noop {} {\bibfield  {journal} {\bibinfo  {journal} {Phys. Rev. D}\
  }\textbf {\bibinfo {volume} {25}},\ \bibinfo {pages} {452} (\bibinfo {year}
  {1982})}\BibitemShut {NoStop}%
\bibitem [{\citenamefont {Birkandan}\ and\ \citenamefont
  {H\c{o}rtacsu}(2017)}]{birkandan_2017}%
  \BibitemOpen
  \bibfield  {author} {\bibinfo {author} {\bibfnamefont {T.}~\bibnamefont
  {Birkandan}}\ and\ \bibinfo {author} {\bibfnamefont {M.}~\bibnamefont
  {H\c{o}rtacsu}},\ }\href@noop {} {\bibfield  {journal} {\bibinfo  {journal}
  {Rep. Math. Phys.}\ }\textbf {\bibinfo {volume} {79}},\ \bibinfo {pages} {81}
  (\bibinfo {year} {2017})}\BibitemShut {NoStop}%
\bibitem [{\citenamefont {Loginov}(2022{\natexlab{a}})}]{loginov_epjc_2022}%
  \BibitemOpen
  \bibfield  {author} {\bibinfo {author} {\bibfnamefont {A.~{\relax Yu}.}\
  \bibnamefont {Loginov}},\ }\href@noop {} {\bibfield  {journal} {\bibinfo
  {journal} {Eur. Phys. J. C}\ }\textbf {\bibinfo {volume} {82}},\ \bibinfo
  {pages} {662} (\bibinfo {year} {2022}{\natexlab{a}})}\BibitemShut {NoStop}%
\bibitem [{\citenamefont {Loginov}(2022{\natexlab{b}})}]{loginov_npb_2022}%
  \BibitemOpen
  \bibfield  {author} {\bibinfo {author} {\bibfnamefont {A.~{\relax Yu}.}\
  \bibnamefont {Loginov}},\ }\href@noop {} {\bibfield  {journal} {\bibinfo
  {journal} {Nucl. Phys. B}\ }\textbf {\bibinfo {volume} {984}},\ \bibinfo
  {pages} {115964} (\bibinfo {year} {2022}{\natexlab{b}})}\BibitemShut
  {NoStop}%
\bibitem [{\citenamefont {Din}\ and\ \citenamefont
  {Zakrzewski}(1980{\natexlab{a}})}]{dz1}%
  \BibitemOpen
  \bibfield  {author} {\bibinfo {author} {\bibfnamefont {A.~M.}\ \bibnamefont
  {Din}}\ and\ \bibinfo {author} {\bibfnamefont {W.~J.}\ \bibnamefont
  {Zakrzewski}},\ }\href@noop {} {\bibfield  {journal} {\bibinfo  {journal}
  {Nucl. Phys. B}\ }\textbf {\bibinfo {volume} {174}},\ \bibinfo {pages} {397}
  (\bibinfo {year} {1980}{\natexlab{a}})}\BibitemShut {NoStop}%
\bibitem [{\citenamefont {Din}\ and\ \citenamefont
  {Zakrzewski}(1980{\natexlab{b}})}]{dz2}%
  \BibitemOpen
  \bibfield  {author} {\bibinfo {author} {\bibfnamefont {A.~M.}\ \bibnamefont
  {Din}}\ and\ \bibinfo {author} {\bibfnamefont {W.~J.}\ \bibnamefont
  {Zakrzewski}},\ }\href@noop {} {\bibfield  {journal} {\bibinfo  {journal}
  {Phys. Lett. B}\ }\textbf {\bibinfo {volume} {95}},\ \bibinfo {pages} {426}
  (\bibinfo {year} {1980}{\natexlab{b}})}\BibitemShut {NoStop}%
\bibitem [{\citenamefont {Din}\ and\ \citenamefont {Zakrzewski}(1981)}]{dz3}%
  \BibitemOpen
  \bibfield  {author} {\bibinfo {author} {\bibfnamefont {A.~M.}\ \bibnamefont
  {Din}}\ and\ \bibinfo {author} {\bibfnamefont {W.~J.}\ \bibnamefont
  {Zakrzewski}},\ }\href@noop {} {\bibfield  {journal} {\bibinfo  {journal}
  {Nucl. Phys. B}\ }\textbf {\bibinfo {volume} {182}},\ \bibinfo {pages} {151}
  (\bibinfo {year} {1981})}\BibitemShut {NoStop}%
\bibitem [{\citenamefont {Slavyanov}\ and\ \citenamefont
  {Lay}(2000)}]{Slavyanov}%
  \BibitemOpen
  \bibfield  {author} {\bibinfo {author} {\bibfnamefont {S.~Y.}\ \bibnamefont
  {Slavyanov}}\ and\ \bibinfo {author} {\bibfnamefont {W.}~\bibnamefont
  {Lay}},\ }\href@noop {} {\emph {\bibinfo {title} {Special Functions: A
  Unified Theory Based on Singularities}}}\ (\bibinfo  {publisher} {Oxford
  University Press},\ \bibinfo {address} {Oxford},\ \bibinfo {year}
  {2000})\BibitemShut {NoStop}%
\bibitem [{\citenamefont {Ronveaux}(1995)}]{Ronveaux}%
  \BibitemOpen
  \bibinfo {editor} {\bibfnamefont {A.}~\bibnamefont {Ronveaux}},\ ed.,\
  \href@noop {} {\emph {\bibinfo {title} {Heun's Differential Equations}}}\
  (\bibinfo  {publisher} {Oxford University Press},\ \bibinfo {address}
  {Oxford},\ \bibinfo {year} {1995})\BibitemShut {NoStop}%
\bibitem [{\citenamefont {Olver}\ \emph {et~al.}(2010)\citenamefont {Olver},
  \citenamefont {Lozier}, \citenamefont {Boisvert},\ and\ \citenamefont
  {Clark}}]{DLMF}%
  \BibitemOpen
  \bibinfo {editor} {\bibfnamefont {F.~W.~J.}\ \bibnamefont {Olver}}, \bibinfo
  {editor} {\bibfnamefont {D.~W.}\ \bibnamefont {Lozier}}, \bibinfo {editor}
  {\bibfnamefont {R.~F.}\ \bibnamefont {Boisvert}},\ and\ \bibinfo {editor}
  {\bibfnamefont {C.~W.}\ \bibnamefont {Clark}},\ eds.,\ \href@noop {} {\emph
  {\bibinfo {title} {NIST Handbook of Mathematical Functions}}}\ (\bibinfo
  {publisher} {Cambridge University Press},\ \bibinfo {address} {Cambridge},\
  \bibinfo {year} {2010})\BibitemShut {NoStop}%
\bibitem [{\citenamefont {Landau}\ and\ \citenamefont
  {Lifshitz}(1977)}]{LandauIII}%
  \BibitemOpen
  \bibfield  {author} {\bibinfo {author} {\bibfnamefont {L.~D.}\ \bibnamefont
  {Landau}}\ and\ \bibinfo {author} {\bibfnamefont {E.~M.}\ \bibnamefont
  {Lifshitz}},\ }\href@noop {} {\emph {\bibinfo {title} {Quantum Mechanics:
  Non-Relativistic Theory. Vol. 3 (3rd ed.)}}}\ (\bibinfo  {publisher}
  {Pergamon Press},\ \bibinfo {address} {Oxford},\ \bibinfo {year}
  {1977})\BibitemShut {NoStop}%
\bibitem [{\citenamefont {Taylor}(1972)}]{Taylor}%
  \BibitemOpen
  \bibfield  {author} {\bibinfo {author} {\bibfnamefont {J.~R.}\ \bibnamefont
  {Taylor}},\ }\href@noop {} {\emph {\bibinfo {title} {Scattering Theory:
  Quantum Theory on Nonrelativistic Collisions}}}\ (\bibinfo  {publisher} {John
  Wiley \& Sons},\ \bibinfo {address} {New York},\ \bibinfo {year}
  {1972})\BibitemShut {NoStop}%
\bibitem [{\citenamefont {Prudnikov}\ \emph {et~al.}(1990)\citenamefont
  {Prudnikov}, \citenamefont {Brychkov},\ and\ \citenamefont {Marichev}}]{PBM}%
  \BibitemOpen
  \bibfield  {author} {\bibinfo {author} {\bibfnamefont {A.}~\bibnamefont
  {Prudnikov}}, \bibinfo {author} {\bibfnamefont {Y.~A.}\ \bibnamefont
  {Brychkov}},\ and\ \bibinfo {author} {\bibfnamefont {O.}~\bibnamefont
  {Marichev}},\ }\href@noop {} {\emph {\bibinfo {title} {Integrals and Series.
  Vol. 3}}}\ (\bibinfo  {publisher} {Gordon and Breach Science Publishers},\
  \bibinfo {address} {New York},\ \bibinfo {year} {1990})\BibitemShut {NoStop}%
\bibitem [{Mat()}]{Mathematica}%
  \BibitemOpen
  \href@noop {} {\bibinfo {title} {Wolfram {R}esearch, {I}nc., {M}athematica,
  {V}ersion 12.2}},\ \bibinfo {note} {{C}hampaign, IL (2020)}\BibitemShut
  {NoStop}%
\bibitem [{Map(2019)}]{Maple}%
  \BibitemOpen
  \href {www.maplesoft.com} {\emph {\bibinfo {title} {Maple User Manual}}},\
  \bibinfo {organization} {Maplesoft},\ \bibinfo {address} {Waterloo, Canada}
  (\bibinfo {year} {2019})\BibitemShut {NoStop}%
\bibitem [{\citenamefont {Levinson}(1949)}]{levinson_49}%
  \BibitemOpen
  \bibfield  {author} {\bibinfo {author} {\bibfnamefont {N.}~\bibnamefont
  {Levinson}},\ }\href@noop {} {\bibfield  {journal} {\bibinfo  {journal}
  {Danske Vidensk. Selsk. K. Mat.-Fys. Medd.}\ }\textbf {\bibinfo {volume}
  {25}},\ \bibinfo {pages} {9} (\bibinfo {year} {1949})}\BibitemShut {NoStop}%
\end{thebibliography}%
\clearpage

\clearpage

\clearpage

\end{document}